\PassOptionsToPackage{hyphens}{url}

\documentclass[camera,hyperref,letterpaper,nomarginnotes,nonarrowgutter]{jpaper}

\usepackage{fancyhdr}

\usepackage[nospace]{cite}

\usepackage{amsfonts}

\usepackage{amssymb}

\usepackage{algorithmic}
\usepackage[linesnumbered]{algorithm2e}
\usepackage{balance}
\usepackage{textcomp}
\usepackage{tabularray}
\usepackage{tabularx}
\usepackage{dblfloatfix}
\usepackage{glossaries}
\usepackage{tablefootnote}
\usepackage[most]{tcolorbox}
\usepackage{eso-pic}

\usepackage{pifont}
\newcommand{\cmark}{\ding{51}}
\newcommand{\xmark}{\ding{55}}

\usepackage{footnote}
\makesavenoteenv{table*}

\definecolor{urlblue}{HTML}{319dd6}
\hypersetup{
    breaklinks=true,
    colorlinks=true,
    linkcolor=blue,
    citecolor=blue,
    urlcolor=urlblue,
 }

\def\UrlBreaks{\do\/\do-\/\do.\/\do:}

\expandafter\def\expandafter\UrlBreaks\expandafter{\UrlBreaks
  \do\a\do\b\do\c\do\d\do\e\do\f\do\g\do\h\do\i\do\j
  \do\k\do\l\do\m\do\n\do\o\do\p\do\q\do\r\do\s\do\t
  \do\u\do\v\do\w\do\x\do\y\do\z\do\A\do\B\do\C\do\D
  \do\E\do\F\do\G\do\H\do\I\do\J\do\K\do\L\do\M\do\N
  \do\O\do\P\do\Q\do\R\do\S\do\T\do\U\do\V\do\W\do\X
  \do\Y\do\Z}

\newcommand{\titlenamefull}{Terracotta}
\newcommand{\titlename}{Terracotta}
\newcommand{\field}[1]{\textit{#1}}

\newcommand{\citescaling}{\cite{mutlu2013memory, mutlu2014research, patel2022case, patel2024rethinking,
    mutlu2017rowhammer
}}

\newcommand{\citebottleneck}{\cite{
    mutlu2013memory, mutlu2014research, mutlu2017rowhammer,
    mutlu2025memory, mutlu2024memory,
    mutlu2022modern, mutlu2025modernprimerprocessingmemory, mutlu2019processing,
    mutlu2019rowhammer, mutlu2019enabling, mutlu2020intelligentdate, mutlu2023fundamentally,
    yuksel2026memory, ghose2019processing,
    patel2022case, patel2024rethinking,
    hong2010memory,
    wilkes2001memory, sigarch1995-wulf-memwall,
    kanev2015profiling, dean2013tail, ferdman2012clearing,
    ghose2018your, vogelsang2010understanding, wang2016reducing, pandiyan2014quantifying,
    ahn2015scalable, ahn2023retrospective, ahn2015pim, koppula2019eden, boroumand2018google,
    seshadri2013rowclone,seshadri2017ambit,
    kim2014flipping, luo2023rowpress, yuksel2025columndisturb, canpolat2025chronus,
    kim2020revisiting, olgun2025variable, luo2026dejavu, frigo2020trrespass,
    yuksel2025pudhammer,
    qureshi2015avatar,liu2013experimental, mutlu2023retrospectiveexperimental,
    kim2012case,
    chang2017thesis, lee2016reducing,
    patel2019understanding
}}

\newcommand{\citedatamovement}{\cite{
    ghose2019processing, kanev2015profiling, dean2013tail, ferdman2012clearing,
    boroumand2018google,
    pandiyan2014quantifying,
    oliveira2021damov,
    ahn2015scalable, ahn2023retrospective
}}

\ExplSyntaxOn
\cs_generate_variant:Nn \seq_gconcat:NNN { NNc }
\seq_new:N \g_tc_alltech_seq
\seq_new:N \g_tc_groupnames_seq
\cs_new_protected:Npn \tc_cite_group:n #1
  { \exp_args:Nx \cite { \seq_use:cn { g_tc_group_#1_seq } { , } } }
\NewDocumentCommand \newcitegroup { m m }
  {
    \seq_new:c { g_tc_group_#1_seq }
    \seq_gset_from_clist:cn { g_tc_group_#1_seq } {#2}
    \seq_gremove_duplicates:c { g_tc_group_#1_seq }
    \cs_new_protected:cpx { cite #1 } { \tc_cite_group:n {#1} }
    \seq_gput_right:Nn \g_tc_groupnames_seq {#1}
    \seq_gconcat:NNc \g_tc_alltech_seq \g_tc_alltech_seq { g_tc_group_#1_seq }
    \seq_gremove_duplicates:N \g_tc_alltech_seq
  }
\NewDocumentCommand \citetechniques { }
  { \exp_args:Nx \cite { \seq_use:Nn \g_tc_alltech_seq { , } } }
\ExplSyntaxOff

\newcitegroup{pum}{
    mutlu2019processing, mutlu2025modernprimerprocessingmemory, mutlu2022modern, mutlu2024memory, mutlu2025memory,
    seshadri2017ambit, seshadri2016buddy, yuksel2024functionally, yuksel2024simultaneous, xin2020elp2im, besta2021sisa,
    li2017drisa, li2018scope, deng2018dracc, angizi2019graphide, oliveira2022accelerating,
    hajinazar2021simdram, kim2019drange, gao2019computedram, gao2022fracdram, olgun2021quac,
    kubo2024bulk, yuksel2024pulsar, nadig2026conduit, bostanci2022drstrange, ali2020inmemory,
    jahshan2024majork, garzon2026cadm, yaron2026bindram, angizi2019redram, zhou2022redlut,
    afifi2024artemis, delima2026count2multiply, li2025pimsum, kubo2025mvdram, kubo2025pudtune,
    zhou2022lt, tokuda2026clutch, kim2018the, baser2026indram, soleimani2025wisedram,
    seshadri2013rowclone, shin2024prada, shin2023thesis, oliveira2024mimdram, ferreira2022pluto,
    deng2019lacc, oliveira2025proteus, seshadri2015fast,
    seshadri2016processing, seshadri2019dram, olgun2022pidram, tokuda2026pudghost
}

\newcitegroup{maintenance}{
    kim2014flipping, kim2020revisiting, frigo2020trrespass, canpolat2025chronus,
    yuksel2025columndisturb, luo2023rowpress, olgun2025variable, yuksel2025pudhammer, luo2026dejavu,
    mutlu2019retrospective, mutlu2023retrospectiveflipping, hassan2021utrr, marazzi2022protrr, brasser2017can, chakraborty2019deep,
    son2017making, gomez2016dram, konoth2018zebram, kim2019effective, wu2019protecting,
    aweke2016anvil, zhang2022softtrr, van2018guardion, loughlin2022moesiprime, enomoto2022efficient,
    ajorpaz2022evax, naseredini2022alarm, joardar2022machine, zhang2020leveraging, kang2020cattwo,
    bennett2021panopticon, yaglikci2021security, han2021surround, saroiu2022configure, park2022row,
    wi2023shadow, bock2019rip, gude2023defending, orosa2021deeper, marazzi2023rega,
    loughlin2021stop, joardar2022learning, taneja2026mirza, qureshi2025moat, taneja2025dream,
    lin2025cncprac, jaleel2024pride, qureshi2025mint, wu2026apt, qureshi2026salt, qureshi2026coda,
    taneja2025roguerfm, qureshi2025autoprac, lang2023blaster, lu2025counterpoint, woo2025qprac,
    wi2026rowarmor, alajmi2026dynamic, wang2026flipshield, qureshi2026elppir, hong2023dsac,
    du2025divide, saxena2025citadel, jattke2025mcsee, bostanci2025revisiting, kim2026sok,
    wang2026scaledisturb, woo2026loaded, kim2026pvac, lee2026beyond, rh-cisco, rh-apple, rh-hp,
    rh-lenovo, ghasempour2015armor, saroiu2022price,
    hassan2024smd, lee2019twice, seyedzadeh2017cbt, canpolat2024understanding, qureshi2022hydra,
    fisch2017dram, wang2019detect, you2019mrloc, juffinger2023csi, manzhosov2022revisiting,
    devaux2021method, fakhrzadehgan2022safeguard, woo2023scalable, kim2023ddr5, wang2021discreet,
    yaglikci2021blockhammer, canpolat2024breakhammer, yaglikci2024spatial, vig2018rapid,
    lee2021cryoguard, kakolyris2026columnkeeper, bostanci2025understanding, saxena2024start,
    saxena2023pt, saxena2024rubix, saxena2024impress, greenfield2012throttling,
    greenfield2014method, bains2016row, qazi2025dram, woo2025dapper, woo2025when,
    tugrul2025understanding, taneja2026from, saroiu2026from,
    bains2015rowref, bains2016distributed, bains15method, hassan2019crow, vittal2025mopac,
    seyedzadeh2018cbt, park2020graphene, kim2022mithril, kim2014architectural, bostanci2024comet,
    olgun2024abacus, saileshwar2022randomized, saxena2022aqua, yaglikci2022hira,
    qureshi2015avatar,
    chang2014improving,
    zhang2014cream, cui2014dtail, riho2014partial, bhati2015flexible,
    liu2012raidr, mutlu2023retrospectiveraidr, nair2014refresh, baek2014refresh, das2018vrl, bhati2013coordinated,
    emma2008rethinking, ghosh2007smart, isen2009eskimo, kim2003block, liu2012flikker,
    mukundan2013understanding, nair2013case, patel2005energy, stuecheli2010elastic, khan2016parbor,
    khan2017detecting, venkatesan2006retention, kim2020charge, nguyen2018nonblocking,
    kwon2021reducing, han2014data, choi2020reducing,
    jung2015omitting, kim2000dynamic, luo2014characterizing, khan2014efficacy, bhati2016dram,
    liu2013experimental, mutlu2023retrospectiveexperimental,
    saleh1990reliability, schroeder2009dram, meza2015revisiting, sridharan2012study,
    sridharan2015memory, rooney2019micron,
    patel2017reaper, patel2021harp, khan2016case, sharifi2017online, gong2018duo, nair2013archshield
}

\newcitegroup{parallel}{
    kim2012case, zhang2014cream, lym2018eruca,
    yaglikci2022hira, chang2014improving, ryu2026piat
}

\newcitegroup{latency}{
    chang2016low,
    hassan2016chargecache, lu2015improving, shin2014nuat, lee2013tiered, lee2017design,
    lee2015adaptive, son2013reducing, kim2018solar, chandrasekar2014exploiting, choi2015multiple,
    lee2024agiledram,
    wang2018reducing, qin2024reducing,
    luo2020clrdram, seongil2014rowbuffer, wang2020figaro, du2024fasa, hassan2019crow
}

\newcitegroup{others}{
    olgun2024sectored, zhang2014half, oconnor2017fine, orosa2021codic
}

\newcommand{\citeramulator}{\cite{ramulator2github, kim2015ramulator, luo2024ramulator2, ramulatorgithub, bostanci2026cleaning, bostanci2026extended, luo2026ramulator2.1,ramulatorsharp}}
\newcommand{\citeddrstd}{\cite{jedecddr, jedec2012ddr3, jedec2020ddr4, jedecddr5, jedecddr5c, jedec2015lpddr4, jedec2020lpddr5, jedec2015hbm, jedec2021hbm}}
\newcommand{\citeddr}{\cite{jedecddr, jedec2012ddr3, jedec2020ddr4, jedecddr5, jedecddr5c, jedec2015lpddr4, jedec2020lpddr5, jedec2015hbm, jedec2021hbm,ramulator2github, kim2015ramulator, luo2024ramulator2, ramulatorgithub, bostanci2026cleaning, bostanci2026extended, luo2026ramulator2.1, ramulatorsharp}}
\newcommand{\citeddrrfu}{\cite{jedecddr, jedec2012ddr3, jedec2020ddr4, jedecddr5, jedecddr5c, jedec2015lpddr4, jedec2020lpddr5, jedec2015hbm, jedec2021hbm}}
\newcommand{\citeweightedspeedup}{\cite{snavely2000symbiotic, eyerman2008systemlevel, michaud2012demystifying}}
\newcommand{\citedrampower}{\cite{chandrasekar2011improved, drampowergithub}}
\newcommand{\citepnm}{\cite{farmahini2015nda, devaux2019true, ke2022near, shin2018mcdram, cho2020mcdramv2, ahn2015scalable, ahn2023retrospective,
    ahn2015pim, besta2021sisa, gao2016hrl,babarinsa2015jafar, kwon202120nm, lee2021hardware,
    kal2023aespa,he2020newton,lee20221ynm,kwon2022system,yoon2025splitsync,yu2025compass,heo2024neupims,
    kim2025pimba,he2025papi,gu2025pim}}

\newcommand{\citeaiml}{\cite{
    boroumand2021google,kwon2022system,he2020newton,lee20221ynm,kwon202120nm,lee2021hardware,
    he2025papi,gu2025pim,kim2025pimba,park2024attacc,heo2024neupims,li2024specpim,nair2026early,
    kal2023aespa,yu2025compass,shin2018mcdram, cho2020mcdramv2
}}

\newcommand{\technewcommands}{\cite{
    seshadri2013rowclone, shin2024prada, shin2023thesis, oliveira2024mimdram, ferreira2022pluto,
    deng2019lacc, oliveira2025proteus, seshadri2015fast,
    bains2015rowref, bains2016distributed, bains15method, hassan2019crow, vittal2025mopac,
    zhang2014cream, cui2014dtail, riho2014partial, bhati2015flexible,
    lym2018eruca, kim2012case,
    wang2020figaro, chang2016low, du2024fasa, wang2018reducing, qin2024reducing, luo2020clrdram,
    seongil2014rowbuffer,
    olgun2024sectored, zhang2014half, oconnor2017fine, orosa2021codic
}}

\newcommand{\techctrlmod}{\cite{
    kim2014flipping, luo2023rowpress, yuksel2025columndisturb, canpolat2025chronus,
    yuksel2025pudhammer,
    qureshi2015avatar,
    yuksel2024functionally, yuksel2024simultaneous, xin2020elp2im, besta2021sisa, seshadri2016buddy,
    li2017drisa, li2018scope, deng2018dracc, angizi2019graphide, oliveira2022accelerating,
    kubo2024bulk, yuksel2024pulsar, nadig2026conduit, bostanci2022drstrange, ali2020inmemory,
    jahshan2024majork, garzon2026cadm, yaron2026bindram, angizi2019redram, zhou2022redlut,
    afifi2024artemis, delima2026count2multiply, li2025pimsum, kubo2025mvdram, kubo2025pudtune,
    zhou2022lt, tokuda2026clutch, kim2018the, baser2026indram, soleimani2025wisedram, olgun2021quac,
    kim2019drange, gao2019computedram, gao2022fracdram, seshadri2017ambit, seshadri2013rowclone,
    shin2024prada, shin2023thesis, oliveira2024mimdram, ferreira2022pluto, deng2019lacc,
    oliveira2025proteus, seshadri2015fast, hajinazar2021simdram,
    hassan2024smd, lee2019twice, seyedzadeh2017cbt, canpolat2024understanding, qureshi2022hydra,
    fisch2017dram, wang2019detect, you2019mrloc, juffinger2023csi, manzhosov2022revisiting,
    devaux2021method, fakhrzadehgan2022safeguard, woo2023scalable, kim2023ddr5, wang2021discreet,
    yaglikci2021blockhammer, canpolat2024breakhammer, yaglikci2024spatial, vig2018rapid,
    lee2021cryoguard, kakolyris2026columnkeeper, bostanci2025understanding, saxena2024start,
    saxena2023pt, saxena2024rubix, saxena2024impress, greenfield2012throttling,
    greenfield2014method, bains2016row, qazi2025dram, woo2025dapper, woo2025when,
    tugrul2025understanding, taneja2026from, saroiu2026from, bains2015rowref, bains2016distributed,
    bains15method, hassan2019crow, vittal2025mopac, seyedzadeh2018cbt, park2020graphene,
    kim2022mithril, kim2014architectural, bostanci2024comet, olgun2024abacus,
    saileshwar2022randomized, saxena2022aqua, yaglikci2022hira,
    chang2014improving, zhang2014cream, cui2014dtail, riho2014partial, bhati2015flexible,
    liu2012raidr, mutlu2023retrospectiveraidr, nair2014refresh, baek2014refresh, das2018vrl, bhati2013coordinated,
    emma2008rethinking, ghosh2007smart, isen2009eskimo, kim2003block, liu2012flikker,
    mukundan2013understanding, nair2013case, patel2005energy, stuecheli2010elastic, khan2016parbor,
    khan2017detecting, venkatesan2006retention, kim2020charge, nguyen2018nonblocking,
    kwon2021reducing, han2014data, choi2020reducing,
    patel2017reaper, sharifi2017online, gong2018duo, nair2013archshield, patel2021harp,
    khan2016case,
    ryu2026piat, lym2018eruca, kim2012case,
    wang2020figaro, chang2016low, du2024fasa, hassan2016chargecache, lu2015improving, shin2014nuat,
    lee2013tiered, lee2017design, lee2015adaptive, kim2018solar, son2013reducing,
    chandrasekar2014exploiting, choi2015multiple, lee2024agiledram, luo2020clrdram,
    seongil2014rowbuffer,
    olgun2024sectored, zhang2014half, oconnor2017fine, orosa2021codic
}}

\definecolor{desertrose}{RGB}{255, 235, 225}
\definecolor{espresso}{rgb}{0.23, 0.15, 0.14}
\definecolor{lavender}{RGB}{250, 242, 250}
\definecolor{casper}{RGB}{222, 229, 232}

\newcommand*\circled[1]{%
    \leavevmode\unskip\nobreak\hspace{0.15em}%
    \tikz[baseline=(char.base)]{%
        \node[shape=circle, draw=espresso, fill=desertrose, minimum size=1em,inner sep=0.5pt] (char) {\textcolor{espresso}{#1}};}}
\newcommand*\tCircled[1]{
    \leavevmode\unskip\nobreak\hspace{0.15em}%
    \tikz[baseline=(char.base)]{
        \node[shape=circle, draw=espresso, fill=desertrose, minimum size=1em,inner sep=0.5pt] (char) {\textcolor{espresso}{#1}};}}
\newcommand*\lCircled[1]{
    \leavevmode\unskip\nobreak\hspace{0.15em}%
    \tikz[baseline=(char.base)]{
        \node[shape=circle, draw=black, fill=lavender, minimum size=1em,inner sep=0.5pt] (char) {\textcolor{black}{#1}};}}
\newcommand*\cCircled[1]{
    \leavevmode\unskip\nobreak\hspace{0.15em}%
    \tikz[baseline=(char.base)]{
        \node[shape=circle, draw=black, fill=casper, minimum size=1em,inner sep=0.5pt] (char) {\textcolor{black}{#1}};}}

\newcommand{\figref}[1]{Fig.~\ref{#1}}
\newcommand{\secref}[1]{§\ref{#1}}
\newcommand{\tabref}[1]{Table~\ref{#1}}

\ExplSyntaxOn
\cs_generate_variant:Nn \seq_gconcat:NNN { NNc }
\seq_new:N \g_tc_analyzed_seq
\seq_new:N \g_tc_ucnames_seq
\int_new:N \g_tc_uccites_int
\cs_new_protected:Npn \tc_cite_uc:n #1
  { \exp_args:Nx \cite { \seq_use:cn { g_tc_uc_#1_seq } { , } } }
\NewDocumentCommand \newusecasegroup { m m }
  {
    \seq_new:c { g_tc_uc_#1_seq }
    \seq_gset_from_clist:cn { g_tc_uc_#1_seq } {#2}
    \seq_gremove_duplicates:c { g_tc_uc_#1_seq }
    \cs_new_protected:cpx { cite #1 } { \tc_cite_uc:n {#1} }
    \seq_gput_right:Nn \g_tc_ucnames_seq {#1}
    \int_gadd:Nn \g_tc_uccites_int { \seq_count:c { g_tc_uc_#1_seq } }
    \seq_gconcat:NNc \g_tc_analyzed_seq \g_tc_analyzed_seq { g_tc_uc_#1_seq }
    \seq_gremove_duplicates:N \g_tc_analyzed_seq
  }
\NewDocumentCommand \citeanalyzedtechniques { }
  { \exp_args:Nx \cite { \seq_use:Nn \g_tc_analyzed_seq { , } } }
\cs_new:Npn \tc_count_text:n #1
  {
    \int_case:nnF {#1}
      {
        {0}{zero} {1}{one} {2}{two} {3}{three} {4}{four}
        {5}{five} {6}{six} {7}{seven} {8}{eight} {9}{nine}
      }
      { \int_eval:n {#1} }
  }
\cs_new:Npn \numanalyzedcategories
  { \tc_count_text:n { \seq_count:N \g_tc_ucnames_seq } }
\cs_new:Npn \numanalyzedtechniquescited
  { \tc_count_text:n { \g_tc_uccites_int } }
\ExplSyntaxOff

\newusecasegroup{pud}{seshadri2017ambit, seshadri2013rowclone, shin2024prada, gao2022fracdram, gao2019computedram}

\newusecasegroup{probrd}{vittal2025mopac, kim2014flipping}

\newusecasegroup{salp}{kim2012case, chang2014improving, yaglikci2022hira}

\newusecasegroup{latred}{chang2016low, wang2020figaro, hassan2019crow, lee2013tiered, hassan2016chargecache,
    du2024fasa, shin2014nuat, lu2015improving}

\newusecasegroup{fga}{oconnor2017fine, zhang2014half, olgun2024sectored}

\newusecasegroup{rng}{kim2019drange, olgun2021quac}

\newusecasegroup{detrd}{kim2014flipping, seyedzadeh2018cbt, park2020graphene, kim2022mithril,
    kim2014architectural, saxena2022aqua, saileshwar2022randomized, olgun2024abacus, bostanci2024comet}

\newcolumntype{Y}[1]{>{\raggedright\arraybackslash}m{\dimexpr #1\textwidth-2\tabcolsep-1.2\arrayrulewidth\relax}}
\newcommand{\usecasestable}{%
\footnotesize
\setlength{\tabcolsep}{3pt}
\renewcommand{\arraystretch}{1.1}
\begin{tabular}{|Y{0.245}||Y{0.200}|Y{0.180}|Y{0.200}|Y{0.175}|}
    \hline
    \rowcolor{desertrose}
    \textcolor{espresso}{\textbf{Category}} & \textcolor{espresso}{\textbf{State}} & \textcolor{espresso}{\textbf{Triggers}} & \textcolor{espresso}{\textbf{Updates}} & \textcolor{espresso}{\textbf{Actions}} \\
    \hline\hline
    \mbox{Processing-using-DRAM} \citepud{}
        & Timers for~custom commands
        & Fire on~a~PuD request (e.g.,~\texttt{COPY})
        & Reload timers when a custom command is issued
        & Multiple row activation and compute commands \\
    \hline
    Probabilistic controller-side read~disturbance~mitigation \citeprobrd{}
        & Single-bit register to~mark a~preventive action
        & Fire when register is~set
        & Probabilistic setting of a register
        & Lower-overhead preventive~actions \\
    \hline
    Subarray-level parallelism \citesalp{}
        & Set of parallelizable requests; Timers for custom commands
        & Fire on parallelizable requests
        & Track parallelizable requests; Reload timers
        & Subarray-level control~commands \\
    \hline
    Access~latency~reduction \citelatred{}
        & Fast-access row table:\newline May require timers
        & Fire on fast-access row table hits
        & Insert/evict fast-access rows; Use lower timing constraints
        & Fast-access custom~commands \\
    \hline
        Deterministic controller-side read disturbance~mitigation \citedetrd{}
        & Activation counters
        & Fire when a~counter exceeds a~threshold
        & Increment activation counters and reset on refresh
        & Lower-overhead preventive~actions \\
    \hline
        In-DRAM true random number generation (RNG) \citerng{}
        & Timers for~custom commands
        & Fire on~an~RNG request
        & Reload timers when~a custom command is~issued
        & --- \\
    \hline
    Fine-grained activation~\citefga{}
        & ---
        & Fire on~activations
        & ---
        & Fine-grained activation~commands \\
    \hline
\end{tabular}
}

\begin{document}

\bstctlcite{IEEEtranBSTCTLterracotta}

\newcommand{\thetitle}{\titlename{}: Enabling the Adoption of New DRAM Techniques\\via a Flexible DRAM Interface and Memory Controller}

\pretitle{\begin{center}\normalfont\fontsize{15}{18}\selectfont\bfseries}
\title{\thetitle}

\newcommand{\ausep}{\hspace{0.5em}}
\author{
    Harsh Songara$^1$ \ausep
    Konstantinos Kanellopoulos$^1$ \ausep
    F. Nisa Bostanc\i$^1$ \ausep
    Konstantinos Marios Sgouras$^1$ \\
    Ataberk Olgun$^1$ \ausep
    \.{I}smail Emir Y{\"u}ksel$^1$ \ausep
    Andreas Kosmas Kakolyris$^1$ \ausep
    A. Giray Ya\u{g}l\i k\c{c}\i$^2$ \ausep
    Onur Mutlu$^{1,3}$ \vspace{0.25em} \\
    \normalsize{
        $^1$ETH Z{\"u}rich \hspace{0.5em} $^2$CISPA \hspace{0.5em} $^3$New York University
    }
}

\AddToShipoutPictureFG*{%
  \AtPageUpperLeft{%
    \put(\LenToUnit{\paperwidth-0.675in},\LenToUnit{-0.70in}){%
      \makebox[0pt][r]{%
        \href{https://www.acm.org/publications/policies/artifact-review-and-badging-current}{\includegraphics[height=0.58in]{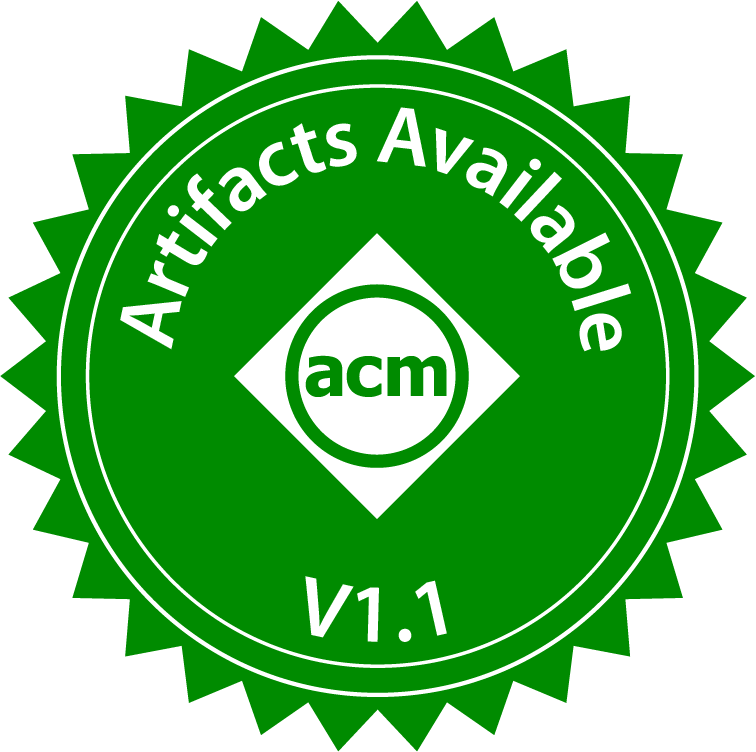}}\hspace{0.06in}%
        \href{https://www.acm.org/publications/policies/artifact-review-and-badging-current}{\includegraphics[height=0.58in]{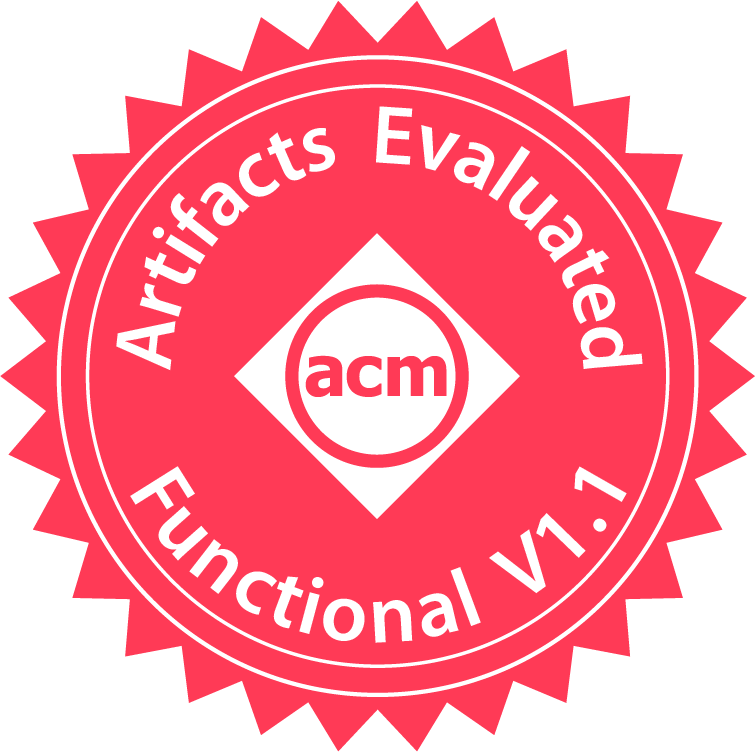}}\hspace{0.06in}%
        \href{https://www.acm.org/publications/policies/artifact-review-and-badging-current}{\includegraphics[height=0.58in]{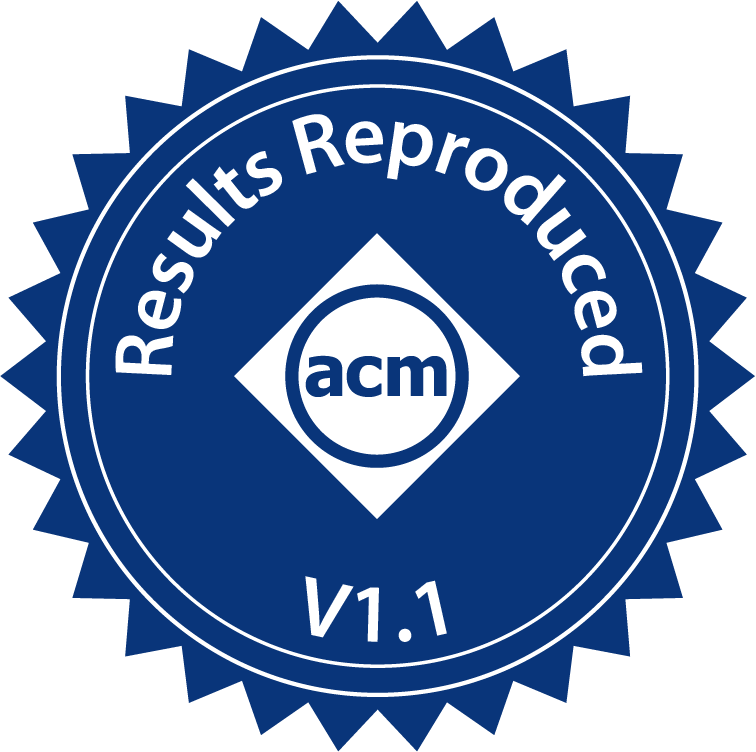}}}}}}
\maketitle

\begin{abstract}

DRAM is the dominant memory technology in modern systems, yet it continues to limit performance, energy efficiency, and system robustness.
To mitigate such limitations, many prior works have proposed novel \emph{DRAM techniques} that change how DRAM is operated to support in-DRAM computation, improve memory access latency and parallelism, and enhance DRAM maintenance and reliability.
However, adopting these techniques one after another in current systems with rigid interfaces and memory controller hardware requires repeated and costly modifications to them,
hindering the deployment of critical performance, energy efficiency, and robustness improvements.

Our goal is to reduce the repeated DRAM interface and memory controller modifications that hinder or slow down the deployment of diverse DRAM techniques.
We observe that the DRAM commands and memory controller structures across several DRAM techniques are \emph{similar}.
Our key idea is to leverage these similarities to compose a set of primitives that DRAM vendors and system designers can use to implement a wide variety of DRAM techniques.
To this end, we propose \emph{\titlename{}}, a new framework that consists of two flexible components:
(i)~\emph{custom command extensions} that enable DRAM vendors to define new commands within a single, standardized interface, and
(ii)~a \emph{programmable memory controller} that system designers can easily and flexibly program to support new DRAM techniques post-silicon fabrication.
Together, these two components enable the deployment of new DRAM techniques via configuration of the memory controller rather than repeated interface and controller modifications.

We design \titlename{} for a DDR5-based system and comprehensively evaluate its performance, energy, and hardware complexity.
We draw three key findings.
First, for four DRAM techniques from four distinct domains including processing-using-DRAM, low-cost DRAM maintenance, subarray-level parallelism, and latency reduction,
\titlename{} retains almost all of the performance benefits (>96\%) that custom implementations provide over a baseline system without these techniques.
Second, a \titlename{}-based composition of two DRAM techniques from the domains of subarray-level parallelism and latency reduction outperforms the \titlename{}-based implementations of the individual techniques,
showing that \titlename{} can enable system improvements by adding techniques without further interface and controller modifications.
Third, \titlename{}'s flexibility incurs low DRAM energy overheads (0.6--3.2\%) and low area and power overheads (0.03\% and 0.56\%, respectively) in a high-end server-grade processor.
\titlename{}'s source code is freely available at \url{https://github.com/CMU-SAFARI/Terracotta}.

\end{abstract}

\section{Introduction}

Dynamic Random Access Memory (DRAM)~\cite{dennard1968field, dennard1974design} is the dominant main memory technology
across a broad range of modern computing systems because of its high capacity and relatively low latency at low cost~\citescaling{}.
However, DRAM continues to limit system performance, energy efficiency, and robustness~\citebottleneck{},
a problem greatly exacerbated today due to the memory demand of emerging Artificial Intelligence (AI) and Machine Learning (ML) workloads~\citeaiml{}.
To mitigate these limitations, many prior works
propose \emph{DRAM techniques} that change how DRAM is architected, operated, and managed in various novel ways~\citetechniques{}.
These techniques
(i)~leverage DRAM's analog processing capability and massive parallelism across columns to enable
processing-using-DRAM (PuD)~\citepum{},
(ii)~enable low-cost maintenance operations (e.g., read disturbance mitigation, refresh, memory scrubbing, and online error profiling)~\citemaintenance{},
(iii)~expose DRAM's parallelism across subarrays to the host system~\citeparallel{}, and
(iv)~reduce {DRAM access latency}~\citelatency{}.

\vspace{2pt}

\noindent\textbf{Motivation.}
DRAM techniques typically require modifications to two parts of the system:
(i)~the DRAM interface~\cite{hassan2024smd}
and (ii)~the memory controller~\cite{zuravleff1997controller, mutlu2007stall, ipek2008self, ipek2023retrospective, mutlu2009parallelism, subramanian2016bliss, mutlu2014research, jacob2010memory}.
First, several DRAM techniques expose novel DRAM functionalities to the memory controller
via new, non-standard DRAM commands~\technewcommands{}, requiring modifications to the DRAM interface.
Second, several DRAM techniques require memory controller hardware support
to determine when to invoke and how to perform a DRAM technique~\techctrlmod{}.

The DRAM interface and controller modifications required by many DRAM techniques pose two deployment obstacles.
First, implementing a DRAM technique that needs DRAM interface and memory controller support may require
the development of a new DRAM standard~\citeddr{}.
This can be prohibitive because it requires agreement among all the stakeholders of the DRAM standard, such as DRAM vendors and system designers.
These stakeholders may have conflicting goals, approaches, and interests~\cite{patel2022case,patel2024rethinking,saroiu2022price,hassan2024smd},
potentially hindering or slowing down the adoption of techniques that benefit a subset of stakeholders.
Second, even if the stakeholders agree to support a DRAM technique in future computing systems,
already-deployed systems \emph{cannot} easily adopt it.
Such systems remain in the field for several years~\cite{siddiqua2017lifetime,saroiu2026from,micronproduct,ramexchangelegacy}
(or even decades for satellite~\cite{aerospacemajority,fox2013satellite}
and space systems~\cite{curiosity,spaceflightcuriosity,berger2001the,3dradiationtolerantsdram,3dradiationtolerantddr2,3dradiationtolerantddr3,
parker,bepicolombo,mars2020,emit,voyager}).
During this time, new DRAM techniques may become valuable and useful to help address the changing memory demands of various workloads~\cite{jouppi2021ten,jouppi2023tpu,boroumand2018google,boroumand2021google,ghose2019processing,kim2025pimba,ahn2015scalable,besta2021sisa,gu2025pim,he2025papi,park2024attacc}, including fast-evolving ones like AI and ML.
Together, these two obstacles hinder the adoption of new DRAM techniques.
Both obstacles stem from the rigidity of current DRAM interfaces and memory controllers, whose supported commands and control logic are fixed at design time.
Adopting techniques that require new commands and controller support therefore entails repeated and costly interface and controller modifications,
depriving computing systems of performance and energy efficiency improvements and essential security features~\cite{patel2024rethinking,hassan2024smd,frigo2020trrespass,jattke2022blacksmith,saroiu2026from,taneja2026from,kim2020revisiting}.

\textbf{Our goal}
is to reduce the repeated interface and controller modifications that hinder or slow down the deployment of diverse DRAM techniques.
Our analysis of \numanalyzedtechniquescited{} DRAM techniques~\citeanalyzedtechniques{} across~\numanalyzedcategories{} different categories
reveals that these techniques rely on similar command structures and memory controller workflows.
Our \textbf{key idea} is to leverage these similarities to compose a set of \emph{primitives}
that DRAM vendors and system designers can independently use to implement a diverse set of DRAM techniques.
To this end, we propose \textbf{\emph{\titlename{}}}, a new memory control framework that consists of two flexible components:
(i)~\emph{custom command extensions} that enable DRAM vendors to independently define new commands within a single standardized interface extension,
requiring \emph{only} a one-time interface change instead of repeated DRAM interface changes,
and (ii)~a \emph{programmable memory controller} that enables system designers to express memory controller support for new DRAM techniques post-silicon fabrication.

\vspace{2pt}

\noindent\textbf{Custom command extensions.}
We observe that DRAM commands of various techniques have a common structure.
We illustrate this using two examples.
First, PRADA~\cite{shin2024prada,shin2023thesis} introduces a partial row activation command, \texttt{A$^*$}, that activates a row without performing sense amplification.
Second, CROW~\cite{hassan2019crow} introduces a row copy command, \texttt{ACT-c}, that copies data from a source row to a destination row in DRAM.
Both \texttt{A$^*$} and \texttt{ACT-c} perform a DRAM operation in a specific DRAM locality (e.g., a specific bank) and operate on one or more operands (e.g., row addresses).
\titlename{} leverages the common structure in such commands to define three primitives:
(i)~\emph{operation type} (e.g., partial row activation or row copy),
(ii)~\emph{target DRAM locality} (e.g., a specific bank), and
(iii)~\emph{payload} (e.g., row addresses).
\titlename{} uses these primitives to extend the DRAM interface through a one-time change that standardizes an encoding for custom commands.
This encoding \emph{only} defines the structural format (i.e., bit positions and widths) as a fixed agreement between system designers and DRAM vendors,
separating structure from the functional meaning of each command.
\titlename{}'s custom command extensions allow DRAM vendors to expose new DRAM functionalities autonomously, without further interface changes or explicit agreement among the stakeholders of the DRAM standard.

\vspace{2pt}

\noindent\textbf{Programmable memory controller.}
Applying the same approach to the memory controller, we identify common controller workflows across different DRAM techniques.
For instance, two existing DRAM techniques,
ChargeCache~\cite{hassan2016chargecache} and CROW~\cite{hassan2019crow}, reduce row activation latency by leveraging high DRAM cell charge in recently precharged and copied rows, respectively.
Despite targeting different row types, both techniques require the memory controller to track eligible rows, check incoming requests against
these rows, and perform
reduced-latency operations when an incoming request targets an eligible row.
\titlename{} decomposes this common structure into four primitives:
\emph{state}, \emph{triggers}, \emph{updates}, and \emph{actions}.
First, \emph{state} is the microarchitectural information the controller maintains for the operation of a DRAM technique (e.g., recently accessed rows).
Second, a \emph{trigger} fires a DRAM technique when a \emph{condition} over the state and an incoming memory request is met (e.g., a read request targets a tracked row).
Third, an \emph{update} maintains the state based on the memory requests issued by the memory controller (e.g., inserting a row into a tracking table on an activation).
Fourth, an \emph{action} issues custom DRAM commands, if a technique needs them, by leveraging \titlename{}'s command extensions.
\titlename{} exposes programmatic control over these primitives, allowing system designers to implement new DRAM techniques post-silicon fabrication without modifying the memory controller.

\vspace{2pt}

\noindent\textbf{Key benefits.}
\titlename{}'s flexible framework allows both DRAM vendors and system designers to enjoy improvements in autonomy and collaboration in four ways.
First, \titlename{} enables DRAM vendors to unilaterally implement DRAM techniques that they (or their collaborators) see fit,
\emph{without} requiring approval from \emph{all} the stakeholders of the DRAM standard.
Second, \titlename{} allows DRAM vendors and system designers to collaborate beyond the DRAM standard to come up with much-needed innovative solutions~\cite{kim2020revisiting,patel2024rethinking,patel2022case,saroiu2022price,mutlu2024memory,mutlu2025memory,mutlu2022modern,mutlu2025modernprimerprocessingmemory}
to mitigate the limitations of modern DRAM-based systems.
Third, both DRAM vendors and system designers can continue to maintain the proprietary nature of their designs.
Fourth, system designers can freely choose among many DRAM chips that can implement different DRAM techniques, as performance, energy efficiency, and security needs change, enabling differentiation across different vendors.
In turn, \titlename{}'s flexibility can foster a more innovative ecosystem, enabling more rapid deployment of new techniques to computing systems, which would benefit end users.

\vspace{2pt}

\noindent\textbf{Evaluation.}
We design \titlename{} for a DDR5-based system~\cite{jedecddr5,jedecddr5c}
and evaluate it in two settings.
First, we evaluate four techniques from four distinct domains:
(i)~PRADA~\cite{shin2024prada,shin2023thesis} (PuD, \secref{sec:processing-using-dram}),
(ii)~MoPAC-C~\cite{vittal2025mopac} (low-cost DRAM maintenance, \secref{sec:maintenance}),
(iii)~MASA~\cite{kim2012case} (subarray-level parallelism, \secref{sec:parallelism}),
and (iv)~ChargeCache~\cite{hassan2016chargecache} (latency reduction, \secref{sec:memory-latency}).
Second, we evaluate a composition of two different DRAM techniques:~MASA~\cite{kim2012case} and ChargeCache~\cite{hassan2016chargecache}.

We compare the performance and energy efficiency of each \titlename{}-based implementation against
(i)~a conventional DRAM system and (ii)~a dedicated custom implementation, using cycle-level simulations with Ramulator~2.0~\citeramulator{}, integrated with DRAMPower~\citedrampower{}.
We make three key observations from our comprehensive evaluation.
First, \titlename{}-based implementations of the four techniques and the composition of ChargeCache and MASA retain over 96\% of their respective custom implementations' performance benefits over the baseline.
Second, the \titlename{}-based composition of ChargeCache and MASA outperforms the individual \titlename{}-based implementations of ChargeCache and MASA, showing that \titlename{}'s flexibility can enable the benefits of multiple techniques without repeated modifications.
Third, \titlename{}'s flexibility incurs a \emph{small} DRAM energy consumption overhead of 3.2\% for PRADA and {<}1\% for MoPAC-C, MASA, and ChargeCache relative to their respective custom implementations.

\titlename{}'s flexibility comes at \emph{low} hardware complexity.
\titlename{}'s custom command encodings require \emph{no} additional pins, leveraging existing reserved command encoding space in the DDR5 standard (\secref{sec:design-extensions}).
We synthesize \titlename{}'s programmable memory controller at the RTL level using the OpenROAD~flow~\cite{openroad} and the Nangate~45nm library~\cite{nangate45}, and scale it to 10nm using DeepScale~\cite{satyabrata2021deepscaletool}.
Our results show that, relative to a high-end server-grade CPU~\cite{intelemeraldrapids}, \titlename{} incurs \emph{only} 0.03\% die area and 0.56\% thermal design power (TDP) overheads on top of a design that provides custom memory controller support for the four DRAM techniques we evaluate.
\titlename{}'s DRAM-side decoding logic adds \emph{only} 5 flip-flops and fewer than 24 gates to the DRAM center stripe.

\vspace{2pt}
\noindent In this work, we make the following contributions:
\begin{itemize}
    \item We propose \textbf{\titlename{}}, a new memory control framework that reduces the repeated modifications needed to adopt diverse DRAM techniques by introducing flexibility into the DRAM interface and memory controller.
    \item We introduce \textbf{custom command extensions} that extend the DRAM interface through a one-time change, decoupling the structural format of new commands from their functional meaning and allowing vendors to define new commands without further interface changes.
    \item We identify four fundamental \textbf{memory controller primitives}: \emph{state, triggers, updates,} and \emph{actions}, that capture the common operational workflows of DRAM techniques. We expose programmatic control over them, allowing system designers to deploy new techniques post-silicon fabrication.
    \item We design \titlename{} for a DDR5-based system and show that it retains almost all of the performance benefits of custom implementations, composes multiple techniques, and incurs low hardware complexity and DRAM energy overheads over custom implementations.
\end{itemize}

We {hope and} expect that \titlename{} will inspire academia and industry to develop and adopt innovative DRAM techniques and optimizations that unlock improvements in system performance, energy efficiency, and robustness.
By helping reduce stagnation in technique adoption caused by prohibitive standardization processes and rigid interfaces and memory controllers, \titlename{} can accelerate progress from new ideas to practical benefits.
To this end, we make \titlename{} freely available at \url{https://github.com/CMU-SAFARI/Terracotta}.

\section{Background}
\label{sec:background}

\subsection{DRAM Organization and Operation}
\label{sec:dram-organization}

\figref{fig:dram-system-organization} shows a DRAM-based computing system.

\begin{figure}[h!]
    \centering
    \includegraphics[width=1.0\linewidth]{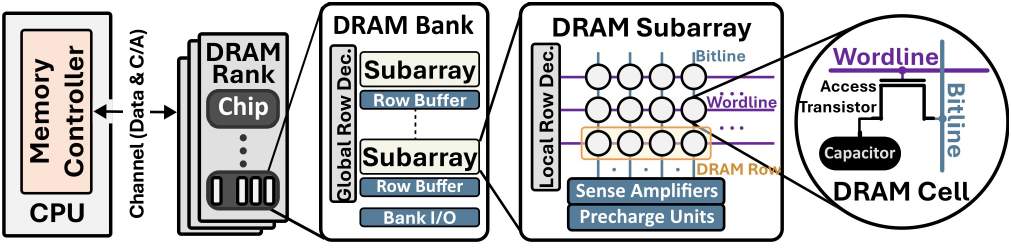}
    \caption{DRAM-based computing system.}
    \label{fig:dram-system-organization}
\end{figure}

\noindent\textbf{DRAM Organization.} DRAM-based systems are typically managed by multiple \emph{Memory Controllers (MCs)}.
Each MC communicates with a \emph{DRAM channel} through an I/O bus that consists of the Command/Address (C/A) bus, to issue commands and addresses, and the data bus, to transfer data.
A DRAM channel consists of one or more \emph{DRAM ranks}, each of which is a group of DRAM chips that operate in lockstep.
A DRAM chip may be divided into multiple \emph{DRAM bank groups}~\cite{jedec2020ddr4,jedecddr5,jedecddr5c,jedec2021hbm} (not shown in \figref{fig:dram-system-organization}).
A DRAM bank group is further divided into multiple \emph{DRAM banks}, each of which is divided into multiple two-dimensional DRAM \emph{subarrays} that share the bank's I/O circuitry.
Within each subarray, DRAM cells are organized into \emph{rows} and \emph{columns}.
A DRAM row consists of DRAM cells that are connected to the same \emph{wordline}.
A DRAM cell connects to a \emph{sense amplifier} via an \emph{access transistor} and a \emph{bitline}, and all sense amplifiers in a subarray form a \emph{row buffer}.
Within each cell, data is stored as \emph{electrical charge} in the \emph{capacitor}.

\noindent\textbf{DRAM Operation.} Reading or updating a DRAM cell requires \emph{activating} (i.e., opening) a DRAM row.
To open a row, the memory controller issues an activate (\texttt{ACT}) command with the bank group, bank, and row addresses.
The \emph{global row decoder} and \emph{local row decoder} of the addressed bank select one of the many wordlines based on the row address.
When the wordline is selected, the cell capacitor connects to the bitline and shares its charge, causing a perturbation in the bitline voltage.
Then, the sense amplifier senses and amplifies the bitline voltage deviation towards V\textsubscript{DD} or 0 V,
fetching the row's data into the {row buffer} and \emph{restoring} the DRAM cell's charge.
The memory controller can then read or modify data in the row buffer using read (\texttt{RD}) and write (\texttt{WR}) commands.
To close a row, the memory controller issues a precharge (\texttt{PRE}) command.
To ensure correctness, the memory controller obeys the DRAM timing constraints~\cite{mutlu2007stall,rixner2000memory,kim2010atlas,mutlu2009parallelism,lee2013tiered,kim2012case,lee2015adaptive,lee2010dram}
specified by the Joint Electron Device Engineering Council (JEDEC)~\citeddr{}, an industry standards body
that includes DRAM vendors and system designers. We discuss four key timing constraints.
First, the latency between an \texttt{ACT} command and subsequent \texttt{RD} and \texttt{WR} commands to the same bank must be at least the activation latency (\texttt{tRCD}).
Second, the latency between consecutive \texttt{RD} commands (and consecutive \texttt{WR} commands) must be at least the column-to-column delay (\texttt{tCCD}).
Third, the latency between a \texttt{PRE} command and a subsequent \texttt{ACT} command to the same bank must be at least the precharge latency (\texttt{tRP}).
Fourth, the latency between consecutive \texttt{ACT} commands to the same bank must be at least the row cycle time (\texttt{tRC}).

\subsection{Memory Controller Organization}
\label{sec:ctlr-organization}

The memory controller translates memory requests into DRAM commands and schedules them onto the DRAM channel.
\figref{fig:controller-organization} shows the organization of a memory controller.

\begin{figure}[h!]
    \centering
    \includegraphics[width=1.0\linewidth]{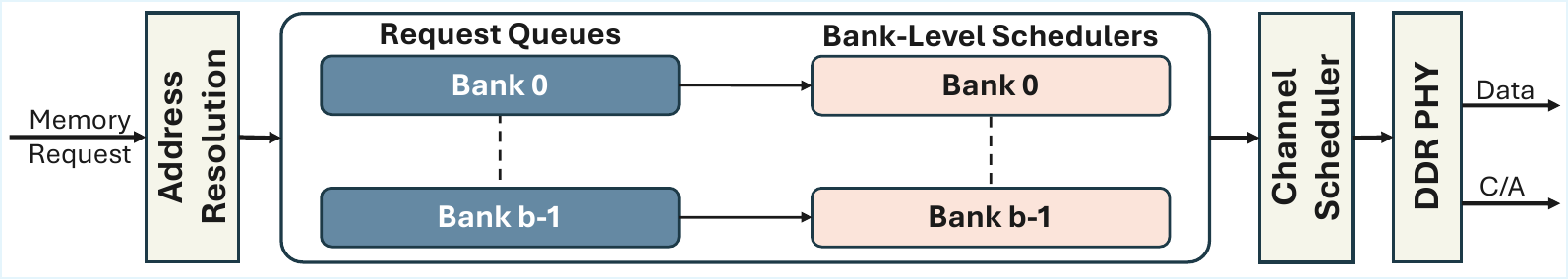}
    \caption{Memory controller organization.}
    \label{fig:controller-organization}
\end{figure}

The memory controller uses \emph{address resolution} to convert each memory request's physical address to its corresponding DRAM address, and places the request into a \emph{request queue}.
The request queue can be organized as multiple bank-partitioned queues or a single unified queue~\cite{mutlu2007stall,rixner2000memory}.
The memory controller then uses a \emph{DRAM scheduler} to decide which command to issue to the DRAM channel.
Typically, the scheduler has two levels~\cite{mutlu2007stall,rixner2000memory}.
The first level consists of \emph{per-bank schedulers}, each of which selects the highest-priority request for its bank and determines the next valid command (e.g.,~\texttt{ACT}, \texttt{RD}, \texttt{PRE}) that the request needs.
The second level is the across-bank \emph{DRAM channel scheduler}, which selects the highest-priority per-bank valid command to issue.
One popular prioritization algorithm employed by DRAM schedulers is FR-FCFS~\cite{rixner2000memory,zuravleff1997controller,mutlu2007stall}.
The \emph{DDR Physical layer (PHY)} converts digital channel-level scheduled commands to analog I/O signals on the DRAM channel.

\section{Motivation \& Goal}
\label{sec:motivation}

Many prior works leverage the internal structure of DRAM to improve the performance, energy efficiency, and robustness of modern systems.
We call such works \emph{DRAM techniques}.
These techniques aim to
(i)~enable in-memory computation by leveraging DRAM's inherently highly parallel internal structure~\citepum{},
(ii)~improve DRAM availability and performance by reducing the overheads of maintenance operations such as refresh, read disturbance mitigation, memory scrubbing, and online error profiling~\citemaintenance{},
(iii)~expose internal DRAM parallelism~\citeparallel{} to the host system,
and (iv)~reduce memory access latency by introducing or identifying fast-access regions~\citelatency{}.

Typically, these techniques rely on support from two parts of the system: (i)~the DRAM interface~\cite{hassan2024smd} and (ii)~the memory controller~\cite{zuravleff1997controller, mutlu2007stall, ipek2008self, ipek2023retrospective, mutlu2009parallelism, subramanian2016bliss, mutlu2014research, jacob2010memory}.
First, various DRAM techniques~\technewcommands{} expose novel DRAM functionalities to the memory controller
by proposing new DRAM commands. These commands are \emph{non-standard} because the existing DRAM interface \emph{only} supports JEDEC-standardized DRAM commands~\citeddr{}.
For example, a recent prior work~\cite{shin2024prada}, PRADA, enables low-cost in-DRAM operations by proposing new DRAM commands. One such command is the \texttt{A$^*$} command which executes only a subset of steps of the standard \texttt{ACT} command: it drives the wordline without performing sense amplification.
Second, many techniques~\techctrlmod{} require dedicated hardware support from the memory controller to determine \emph{when} to invoke and \emph{how} to perform them.
For instance, ChargeCache~\cite{hassan2016chargecache}, a memory~latency~reduction technique, introduces a metadata table to track recently precharged rows
and applies lower timing constraints when accessing such recently precharged rows.

The support that DRAM techniques require from the DRAM interface and controller poses two deployment obstacles.
First, a DRAM technique that needs a new DRAM command requires the development of a new DRAM standard or a revision of an existing standard~\citeddr{}.
Such a change can be prohibitive because it requires agreement among \emph{all} the stakeholders of the DRAM standard (e.g., DRAM vendors and system designers).
These stakeholders may have conflicting goals, approaches, and interests~\cite{patel2022case,patel2024rethinking,saroiu2022price,hassan2024smd},
potentially hindering the adoption of a technique that benefits a subset of the stakeholders.
Second, even if the stakeholders of the standard agree on a technique, already-deployed computing systems \emph{cannot} easily adopt the technique,
since such systems remain in the field for several years~\cite{siddiqua2017lifetime,saroiu2026from,micronproduct,ramexchangelegacy}.
This problem is particularly acute for remote deployments such as satellite~\cite{aerospacemajority,fox2013satellite}
and space systems~\cite{curiosity,spaceflightcuriosity,berger2001the,3dradiationtolerantsdram,3dradiationtolerantddr2,3dradiationtolerantddr3,parker,bepicolombo,mars2020,emit,voyager} that \emph{cannot} be easily upgraded
and may remain in the field for decades.
Over these long deployment lifetimes, already-deployed systems may benefit from new DRAM techniques to meet the fast-evolving memory demands of various workloads (e.g., AI and ML)~\cite{jouppi2021ten,jouppi2023tpu,boroumand2018google,boroumand2021google,ghose2019processing,kim2025pimba,ahn2015scalable,besta2021sisa,gu2025pim,he2025papi,park2024attacc}.
Yet these systems cannot adopt such techniques.

The DRAM interface and memory controller of current computing systems are \emph{rigid}: they support \emph{only} standard DRAM commands and control logic that is defined at design time.
This design philosophy hinders the deployment of DRAM techniques in two ways.
First, already-deployed systems \emph{cannot} adopt a new DRAM technique, even after the technique is standardized.
Second, future systems can adopt a technique \emph{only} by modifying the DRAM interface and redesigning the memory controller.
Adopting $N$ DRAM techniques one after another therefore requires up to $N$ interface changes and $N$ controller redesigns.
These \emph{repeated} modifications deprive computing systems of performance improvements, energy efficiency gains, and essential security features~\cite{patel2024rethinking,hassan2024smd,frigo2020trrespass,jattke2022blacksmith,saroiu2026from,taneja2026from},
motivating the need for a \emph{less rigid} memory system.

\vspace{2pt}
\emph{\textbf{Our goal}} is to reduce the repeated DRAM interface and memory controller modifications that hinder or slow down the deployment of {diverse} DRAM techniques.
To this end, the memory system needs to be able to
(i)~support new DRAM commands \emph{without} a DRAM interface modification for each new command, and
(ii)~implement memory controller support for new DRAM techniques even \emph{after} the controller is fabricated.

\par
\begin{table*}[!b]
    \small
    \centering
    \caption{\numanalyzedtechniquescited{} DRAM techniques across \numanalyzedcategories{} categories and how they map to \titlename{}.}
    \label{tab:techniques}
    \usecasestable{}
\end{table*}

\section{\titlename{}}
\label{sec:terracotta}

Our analysis of~\numanalyzedtechniquescited{} DRAM techniques~\citeanalyzedtechniques{} across~\numanalyzedcategories{} different categories
reveals that these techniques share similar structures in how they define new commands and how they use the memory controller.
Our \textbf{\emph{key~idea}} is to leverage these similarities to compose a set of configurable \emph{primitives} that DRAM vendors and system designers can use to implement many diverse DRAM techniques.
To this end, we propose \emph{\textbf{\titlename{}}}, a new memory control framework that consists of two flexible components:
(i)~custom command extensions that enable DRAM vendors to define new commands within a single, standardized interface extension (\secref{sec:extensions}),
and (ii)~a programmable memory controller that enables system designers to express controller support for new techniques post-silicon fabrication (\secref{sec:abstractions}).

\subsection{Custom Command Extensions}
\label{sec:extensions}

We observe that DRAM commands from various techniques can be decomposed into three primitives:
(i)~\emph{operation type},
(ii)~\emph{target DRAM locality}, and
(iii)~\emph{payload}.
We illustrate these using PRADA's~\cite{shin2024prada,shin2023thesis} non-standard activation command, \texttt{A$^*$}.
The {operation type} specifies the operation DRAM must perform. For \texttt{A$^*$}, the operation drives a wordline without sense amplification.
The {target DRAM locality} identifies the portion of DRAM targeted by the operation. For \texttt{A$^*$}, the target DRAM locality is the target bank.
The {payload} specifies the operation's operands, such as row, column, or subarray addresses within the target DRAM locality.
For \texttt{A$^*$}, the payload is the row address corresponding to the target wordline.

The existing DRAM interface standardizes both the \emph{command structure},
which specifies the bit positions and widths of the fields encoding these three primitives,
and their \emph{functional meaning}, which specifies how DRAM interprets the values in those fields.
Thus, adding a new command to the DRAM interface requires standardizing both its structure and meaning.
To avoid this, \titlename{} augments the existing interface through a \emph{one-time} change that standardizes only the structure of custom commands,
leaving DRAM vendors to independently define each primitive's functional meaning.
For example, a vendor can define PRADA's \texttt{A$^*$} by assigning an operation type value to "activation without sense amplification" and interpreting the locality and payload fields as the target bank and row address, respectively.
This separation allows DRAM vendors to independently define custom commands within a common command structure \emph{without} further interface standardization.

\subsection{Programmable Memory Controller}
\label{sec:abstractions}

Many DRAM techniques share common memory controller structures and operational workflows despite proposing dedicated hardware.
We formulate these commonalities into four fundamental primitives: \emph{state}, which provides microarchitectural information required by techniques, and \emph{triggers}, \emph{updates}, and \emph{actions}, which form a workflow that executes on every scheduled command.
\figref{fig:prims-controller} shows these primitives, their responsibilities, and their interactions.

\begin{figure}[h!]
    \centering
    \includegraphics[width=1.0\linewidth]{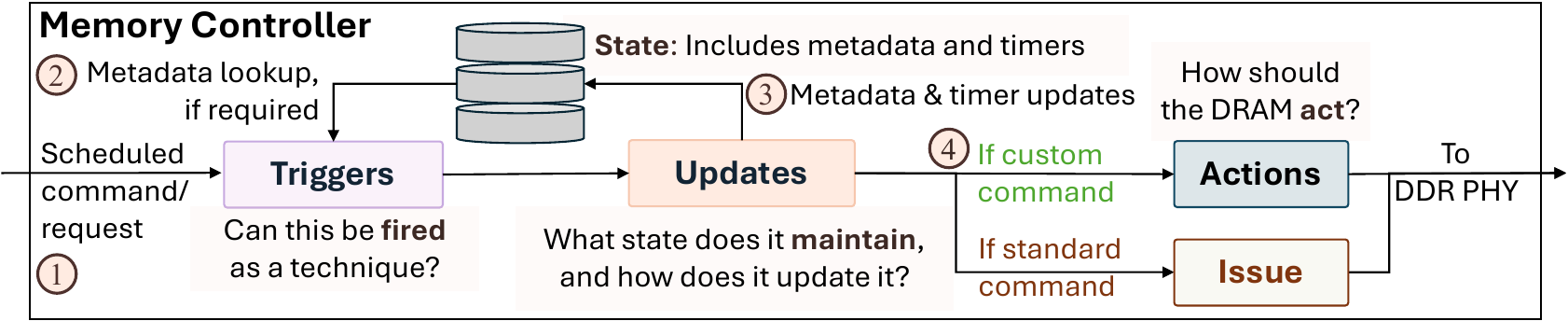}
    \caption{\titlename{}'s memory controller primitives, their responsibilities, and their interactions.}
    \label{fig:prims-controller}
\end{figure}

\begin{figure*}[!t]
    \centering
    \includegraphics[width=1.0\textwidth]{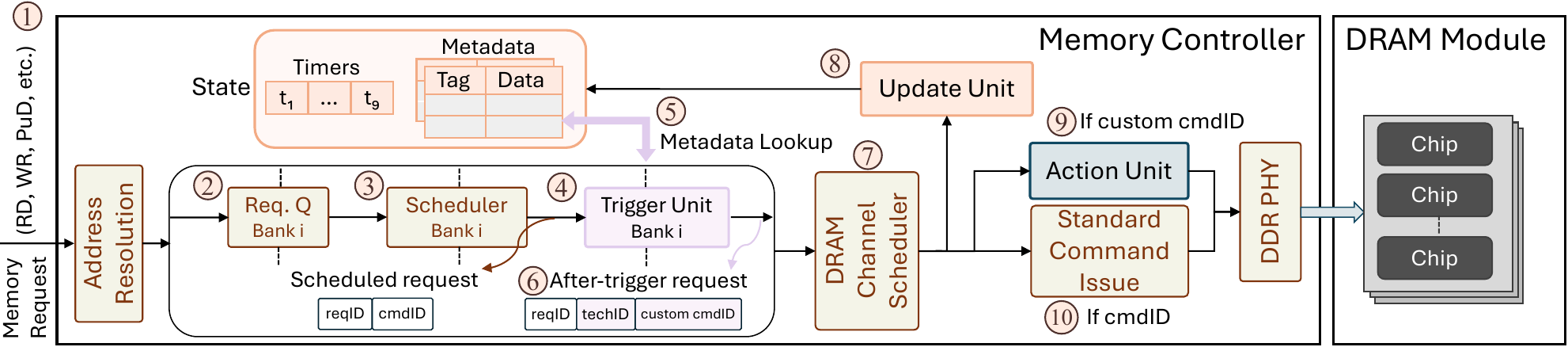}
    \caption{\titlename{}: end-to-end memory request flow.}
    \label{fig:overview}
\end{figure*}

{\emph{State}}
is the {microarchitectural information} the memory controller maintains to execute DRAM techniques.
It consists of:
(i)~\emph{metadata}, which tracks information about DRAM operations (e.g., access history, counters),
and (ii)~\emph{timers}, which ensure that DRAM commands meet timing constraints.
For example, ChargeCache~\cite{hassan2016chargecache} maintains metadata to track recently precharged rows,
and PRADA~\cite{shin2024prada,shin2023thesis} requires timers to enforce the timing constraints of its non-standard commands.

When a command or request is scheduled~\circled{1}, \emph{triggers}
evaluate \emph{conditions} to determine whether a technique should fire.
A condition typically involves:
(i)~the \textit{scheduled command or request} type,
and (ii)~a \textit{metadata} lookup~\circled{2} from state.
For example, ChargeCache~\cite{hassan2016chargecache} fires when a read (\texttt{RD}) request targets a row present in its metadata table of recently precharged rows.
Irrespective of whether a trigger fires, {\emph{updates}}
perform the technique's \emph{state maintenance} operations.
This includes:
(i)~maintaining metadata consulted by triggers,
and (ii)~updating timers to ensure that subsequent commands satisfy timing constraints~\circled{3}.
For example, ChargeCache performs two updates:
(i)~inserting precharged rows into a metadata table,
and (ii)~recording timing constraints for future accesses to such rows.
If a technique requires a non-standard DRAM operation~\circled{4}, the controller provides support for custom commands via {\emph{actions}}.
For example, PRADA's action issues a custom \texttt{A$^*$} command to drive a wordline without sense amplification.

Since these primitives capture common controller structures and are \emph{not} tied to specific functionalities, \titlename{} can expose them as configurable building blocks rather than designing separate fixed logic blocks for each technique.
Doing so allows system designers to \emph{program} state, trigger conditions, update rules, and actions to implement new DRAM techniques, reducing the need to modify the memory controller.
We show next that \titlename{}'s primitives cover a wide range of DRAM techniques.

\subsection{Applicability Across Techniques}
\label{sec:applicability}

\tabref{tab:techniques}
maps the \numanalyzedtechniquescited{} DRAM techniques we analyze to \titlename{}.
Each row breaks down a category by its required state, triggers, updates, and actions (and custom commands).\footnote{Custom commands are tied to actions because an action provides the controller-side logic to issue the custom command.}
While these techniques serve diverse goals, they rely on common underlying structures that \titlename{} captures through its command extensions (\secref{sec:extensions}) and controller primitives (\secref{sec:abstractions}).\footnote{%
Some DRAM techniques propose to also modify the DRAM chip.
Of the~\numanalyzedtechniquescited{} techniques in \tabref{tab:techniques}, 15 require DRAM chip modifications.
\titlename{} provides the DRAM interface and memory controller support required by \emph{all}~\numanalyzedtechniquescited{} techniques, and deliberately leaves the DRAM chip modifications to the DRAM vendors.
We discuss this further in \secref{sec:limitations}.}

\section{\titlenamefull{}: Design Overview}
\label{sec:design}

\figref{fig:overview} illustrates the end-to-end flow of memory requests at runtime through \titlename{}'s memory controller.
The memory controller receives memory requests~\tCircled{1} and,
similar to conventional memory controllers (\secref{sec:background}), places them into per-bank request queues~\tCircled{2}.
The per-bank scheduler~\tCircled{3} selects a request and passes it to the corresponding trigger unit~\tCircled{4}.
The trigger unit evaluates the request against a set of conditions using request attributes such as request ID (\field{reqID}), command ID (\field{cmdID}), and DRAM addresses.
For techniques that depend on metadata, the trigger unit performs a metadata lookup~\tCircled{5}.
If the request satisfies the trigger conditions, the trigger unit \emph{marks} it as eligible for the corresponding technique~\tCircled{6} in two steps.
First, it adds a technique ID (\field{techID}) to identify the triggered technique.
Second, if the technique requires a custom command, the trigger unit replaces the request's \field{cmdID} with the corresponding custom command identifier, \field{custom cmdID}.
The trigger unit then forwards both marked and unmarked requests to the channel scheduler~\tCircled{7}.
The channel scheduler selects a request and simultaneously forwards it to the update unit and the action unit.
The update unit manages memory controller state~\tCircled{8}, such as metadata and timers, based on the scheduled command's \field{techID} and \field{cmdID}.
If the selected command is a custom command~\tCircled{9},
the action unit uses its \field{cmdID} to issue the corresponding custom command to the DDR PHY according to \titlename{}'s custom command extension format.
Otherwise, if the selected command is a standard command~\tCircled{10}, \titlename{} issues it to the DDR PHY through the standard command issue path.

\section{\titlenamefull{}: Implementation}
\label{sec:implementation}

A key challenge in designing \titlename{} is balancing the flexibility needed to support diverse DRAM techniques with the hardware complexity required to provide that flexibility.
This trade-off raises three key implementation questions, each of which exposes a corresponding design space.

\vspace{1pt}
\noindent\textbf{How should custom commands be encoded?}
Several DRAM standards reserve command encoding space for future use~\citeddrrfu{}.
A design that uses this encoding space can support custom commands without adding pins to the C/A bus, but is limited by the size of the reserved space.
Alternatively, a design can add pins to the C/A bus to provide a larger encoding space, at the cost of additional area.

\vspace{1pt}
\noindent\textbf{How should controller state be managed across techniques and banks?}
One design choice is to share the controller state across techniques and banks.
However, this may lead to contention for the shared state, and leave performance on the table in the absence of an intelligent prioritization and fairness scheme.
Alternatively, a design can partition the state across techniques and banks, avoiding contention but potentially leaving some state unused.

\vspace{1pt}
\noindent\textbf{How expressive should the trigger, update, and action logic be?}
The design space spans increasing flexibility, from programmable tables to Coarse-Grained Reconfigurable Arrays~\cite{karunaratne2017hycube,wang2019hycube,kim2012ulp,whatmough2021smiv,prabhakar2017plasticine}, embedded Field Programmable Gate Arrays~\cite{kumar2021post,li2023duet}, and in-order cores~\cite{riscvmini-github}.
With greater flexibility, the implementation of more complex, future techniques may be easier but at the cost of increased hardware complexity.

\vspace{1pt}
Next, we present a concrete implementation of \titlename{} for a system based on DDR5~\cite{jedecddr5,jedecddr5c}, a representative modern DRAM specification.
Our implementation uses DDR5's reserved command encoding space and programmable tables to maintain low hardware complexity, while partitioning state across techniques and banks to maximize their benefits.
We describe the key components of \titlename{}'s implementation in detail, including:
(i)~the custom command encoding (\secref{sec:design-extensions}),
(ii)~the programmable controller microarchitecture~(\secref{sec:design-mc}), and
(iii)~boot-time system support for configuring DRAM techniques (\secref{sec:system-support}).
We conclude the detailed design with a discussion on how \titlename{}'s programmable memory controller does \emph{not} expand the system's attack surface (\secref{sec:security}) and of \titlename{}'s hardware complexity (\secref{sec:overheads}).

\subsection{Custom Command Encoding}
\label{sec:design-extensions}

\figref{fig:design-extensions}
shows the command encodings \titlename{} uses for custom DRAM commands.
The DDR5 C/A bus reserves command encoding space~\cite{jedecddr5,jedecddr5c}, identified by the \texttt{01} pattern in the two most significant bits~\tCircled{1}.
\titlename{} leverages this reserved space for its custom command encoding.
The encoding comprises (i)~a \emph{header} and (ii)~a \emph{payload}, each occupying one C/A bus cycle.
Each custom command consists of a header followed by one or more payloads, one for each operation operand.

\begin{figure}[h!]
    \centering
    \includegraphics[width=1.0\linewidth]{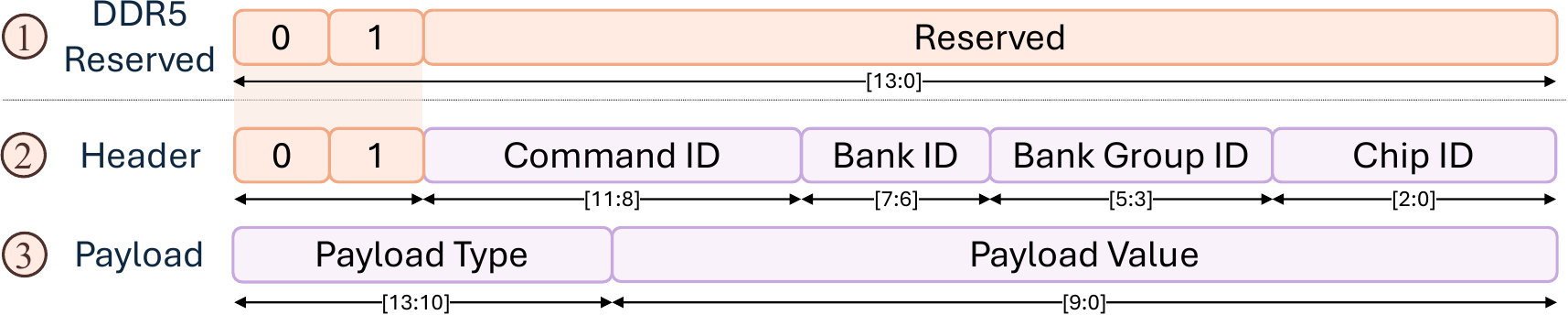}
    \caption{Custom-command structural encoding.}
    \label{fig:design-extensions}
\end{figure}

\noindent\textbf{Header}\tCircled{2}\textbf{.}
The header uses the reserved \texttt{01} pattern to signal a custom command to the DRAM. The header serves three purposes.
First, it conveys the operation type using a 4-bit command ID, allowing a DRAM vendor to define up to 16 different custom commands for each DRAM chip design.\footnote{DDR5~\cite{jedecddr5,jedecddr5c} also provides 4 reserved-for-future-use (RFU) commands that can further expand the encoding space for custom commands.}
Second, because each operation type is associated with a fixed number of payloads, the command ID field of the header \emph{implicitly} determines the number of payloads that follow.
Third, the header encodes the target DRAM locality using bank ID and bank group ID fields.\footnote{The header also carries the Chip ID (CID) field to support 3D-stacked DRAM configurations, following the standard DDR5 command structure.}

\vspace{1pt}
\noindent\textbf{Payload}\tCircled{3}\textbf{.}
A custom command may require multiple \emph{different} operands, such as row and subarray addresses.
Thus, each payload associates a payload value with a 4-bit payload type identifier, allowing a DRAM vendor to define up to 16 different payload types per custom command.

\subsection{Programmable Controller Microarchitecture}
\label{sec:design-mc}

In this subsection, we describe the microarchitectural implementation of \titlename{}'s memory controller primitives.\footnote{We assume component sizes sufficient to support up to four techniques concurrently. Supporting more techniques requires scaling these components and does \emph{not} require changes to the command encodings or controller logic.}

\subsubsection{State: Metadata and Timers}
\label{sec:design-state}

Our implementation partitions \titlename{}'s memory controller state across techniques and DRAM banks to mitigate resource contention,
avoiding the need for a fairness scheme (\secref{sec:implementation}).
The per-bank state comprises two hardware components:
(i)~\emph{metadata tables} and (ii)~\emph{timers}.
First, the metadata table is a 64-entry, 2-way set-associative structure.
Each entry consists of three fields: tag, data, and replacement bits.
To prevent stale data, we employ a configurable age-based entry invalidation policy.
Second, the timers consist of 16 programmable down-counters that enforce the timing constraints introduced by custom commands.
We note that the metadata table capacity (64 entries) and the number of down-counters (16) are \emph{design-time parameters}, analogous to sizing a processor cache.
Scaling these structures to accommodate techniques with larger state requirements does \emph{not} require changes to \titlename{}'s programmable logic (triggers, updates, and actions) or custom command encoding.

\subsubsection{Trigger Unit}
\label{sec:design-triggers}

A trigger serves three functions.
First, \emph{request evaluation} checks each request selected by a bank-level scheduler against the trigger conditions of the configured techniques
to determine the set of eligible techniques.
Second, \emph{C/A bus filtering} excludes some eligible techniques to mitigate critical request starvation, since custom commands may occupy multiple C/A bus cycles (\secref{sec:design-extensions}).
Third, \emph{technique prioritization} selects one technique according to a programmable priority order.
\figref{fig:triggers} illustrates the trigger unit microarchitecture.

\begin{figure}[h!]
    \centering
    \includegraphics[width=1.0\linewidth]{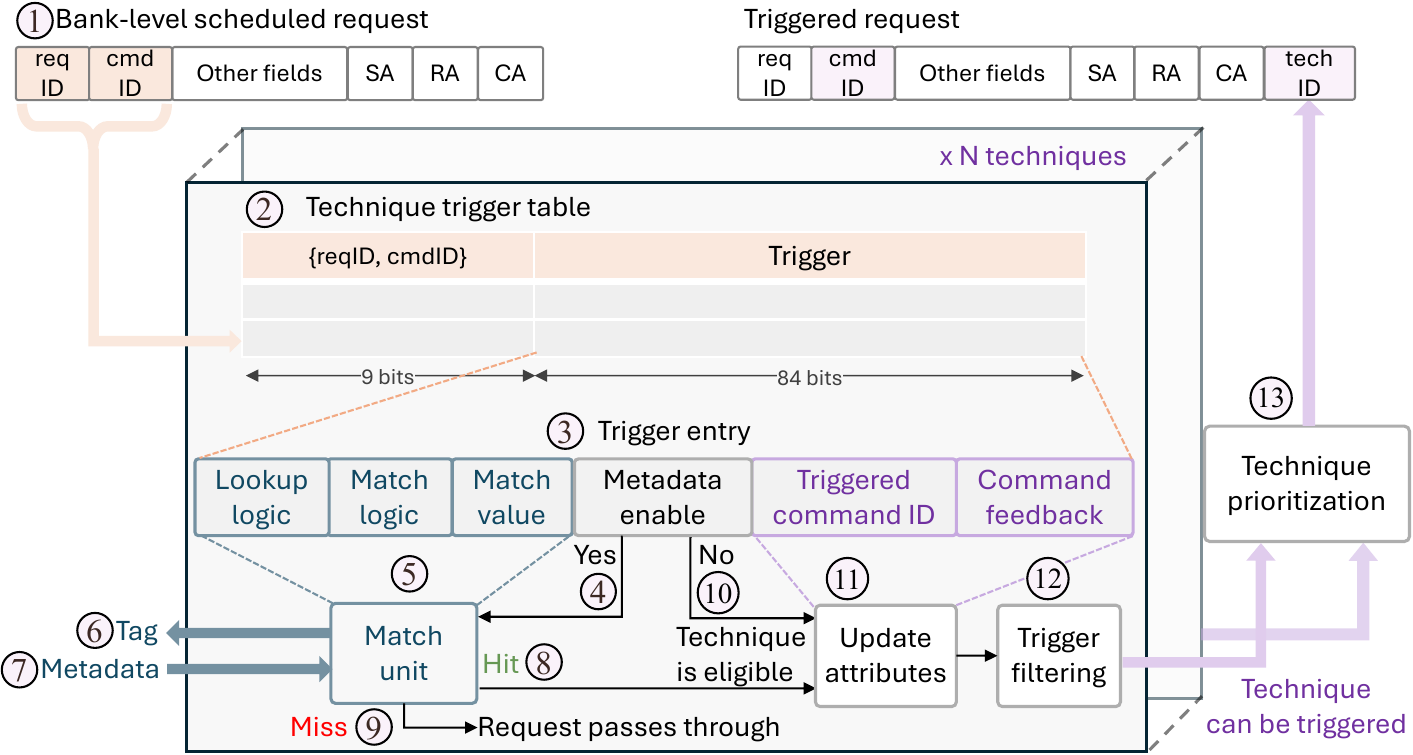}
    \caption{Microarchitecture of the trigger unit.}
    \label{fig:triggers}
\end{figure}

\noindent The trigger unit is partitioned across techniques to evaluate multiple techniques against their trigger conditions in parallel.

\vspace{1pt}
\noindent\textbf{Request evaluation.}
The \field{reqID} and \field{cmdID} fields of the bank-level scheduled request~\lCircled{1}
form the tag used to search each technique's \emph{technique trigger table}\lCircled{2}.
The trigger table is a {programmable}, fully associative structure containing 16 \emph{trigger entries}.
The table returns a {trigger entry}~\lCircled{3} on a hit. The entry consists of six fields:
(i)~metadata enable flag,
(ii)~lookup logic,
(iii)~match logic,
(iv)~match value,
(v)~triggered command ID,
and (vi)~command feedback.
For metadata-dependent techniques, the \emph{metadata enable} flag of the trigger entry is set~\lCircled{4}.
The trigger unit evaluates a metadata match using the \emph{match unit}~\lCircled{5} in two steps.
First, it generates a lookup tag by concatenating up to three DRAM address fields of the request, i.e., subarray address (SA), row address (RA), and column address (CA), as configured by the \emph{lookup logic} field, and uses this tag to look up the technique's bank-partitioned metadata table~\lCircled{6} (\secref{sec:design-state}).
Second, on a metadata table hit, the match unit evaluates the retrieved metadata~\lCircled{7} against the \emph{match value}
using either a direct comparison or a constant modulo operation, as determined by the \emph{match logic} field.
On a metadata match~\lCircled{8}, the technique becomes \emph{eligible}. Otherwise, the request proceeds with its original standard command~\lCircled{9}.
For metadata-independent techniques~\lCircled{10}, a technique trigger table hit alone makes the technique eligible.
For each eligible technique, the trigger unit updates the request's attributes~\lCircled{11} in two ways.
First, if the technique uses a custom command, the trigger unit replaces the request's \field{cmdID} with the \emph{triggered~command~ID} specified in the trigger entry.
Second, the trigger unit provides \emph{command~feedback} to the scheduler, which is either the command that must follow the custom command (e.g., the read after a custom activate) or an indication that the custom command has completed the request.

\vspace{1pt}
\noindent\textbf{C/A bus filtering.}
Some techniques require custom commands that occupy multiple C/A bus cycles (\secref{sec:design-extensions}).
Such techniques can hog the C/A bus, potentially starving critical requests that do not use custom commands, degrading system performance.
To mitigate this issue, the trigger unit evaluates whether an eligible technique should be filtered out based on C/A bus utilization~\lCircled{12}.
The trigger unit employs a channel-level sliding-window counter that tracks the fraction of C/A bus cycles occupied by commands over a configurable window.
When this fraction exceeds a programmable per-technique threshold, the technique is filtered out and the request proceeds with its original standard command.\footnote{The filter can be disabled per technique, ensuring that security-critical and execution-critical (e.g., PuD) techniques always fire when triggered.}

\vspace{1pt}
\noindent\textbf{Technique prioritization.}
If multiple techniques remain eligible for a request, the trigger unit selects one based on a programmable priority order~\lCircled{13},
and marks the request with the selected technique's \field{techID}.
This is useful in prioritizing security-critical techniques.

\vspace{2pt}
\noindent\textbf{Example:} For ChargeCache~\cite{hassan2016chargecache}, the trigger is metadata-dependent.
On a scheduled \texttt{ACT} command, the lookup logic concatenates the subarray address (SA) and row address (RA) to look up the metadata table, which stores recently precharged rows.
A hit indicates the row is still highly charged.
The match logic compares the entry against the match value to confirm its validity.
If the entry is valid, the trigger unit marks the request with the \field{techID} to indicate a ChargeCache hit and provides command feedback to the scheduler.

\subsubsection{Update Unit}
\label{sec:design-updates}
An update maintains the memory controller state based on channel-level scheduled commands.
It serves two functions.
First, \emph{metadata update} modifies the data stored in the technique's metadata table.
Second, \emph{timer update} loads timers with its technique-specific timing constraints.
\figref{fig:updates} illustrates the update unit microarchitecture.

\begin{figure}[h!]
    \centering
    \includegraphics[width=1.0\linewidth]{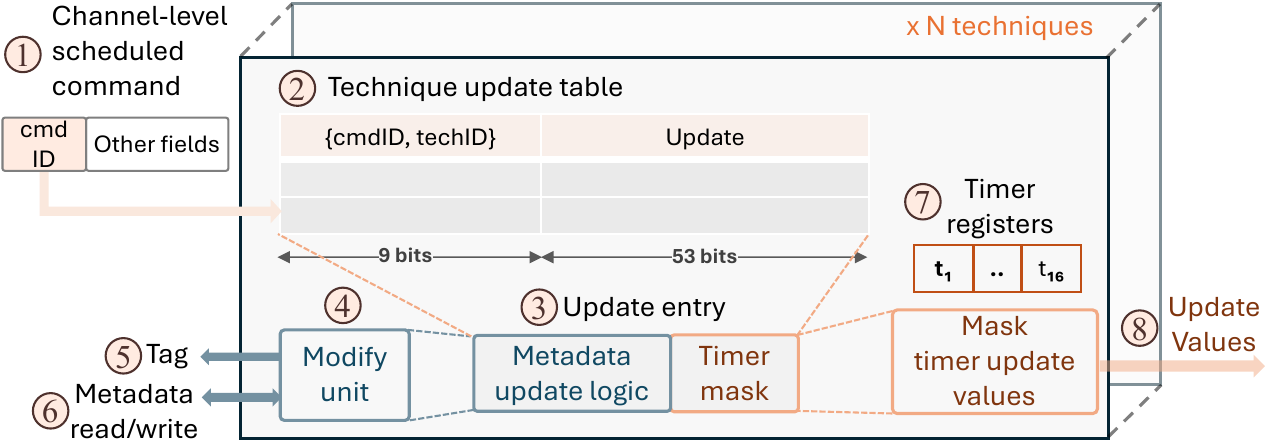}
    \caption{Microarchitecture of the update unit.}
    \label{fig:updates}
\end{figure}

\noindent The update unit is partitioned across techniques so that all per-technique updates are performed in parallel.
For every channel-level scheduled command~\tCircled{1}, the update unit uses the \field{cmdID} and \field{techID} fields to form the tag used to search each technique’s \emph{technique update table}~\tCircled{2}.
Each technique update table is a programmable, fully associative structure containing 8 \emph{update entries}.
On a hit, the update unit performs technique-specific state updates based on the update entry’s two fields~\tCircled{3}:
(i)~metadata update logic, and (ii)~timer mask.

\vspace{1pt}
\noindent\textbf{Metadata update.}
For techniques that maintain metadata, the update unit uses the \emph{metadata update logic} field to configure an arithmetic unit called the \emph{modify unit}~\tCircled{4}.
The {modify unit} performs a read-modify-write operation in two steps.
First, it generates a lookup tag like the trigger lookup logic (\secref{sec:design-triggers}) and uses this tag to look up the technique's bank-partitioned metadata table~\tCircled{5}.
Second, it modifies the retrieved data using one of three operations: adding a constant, setting it to a specified value, or setting it to a specified value with a specified probability.
The modify unit then writes the result back to the metadata table~\tCircled{6}.

\vspace{1pt}
\noindent\textbf{Timer update.}
The update unit employs dedicated per-technique \emph{timer registers}~\tCircled{7} to store timer update values for techniques that introduce new timing constraints or modify existing ones.
The \emph{timer mask} field of the update entry selects which of the bank's timers (\secref{sec:design-state}) are loaded with these values.

\vspace{2pt}
\noindent\textbf{Example:} ChargeCache~\cite{hassan2016chargecache} performs two updates.
First, on a \texttt{PRE} command, the update unit inserts the currently open row into the metadata table to track highly charged rows.
Second, on a ChargeCache \texttt{ACT} command, the updates unit applies reduced timing constraints that are stored in the timer registers.

\subsubsection{Action Unit}
\label{sec:design-actions}

An action enables the memory controller to issue a custom command over the C/A bus.
Each custom command consists of two parts: \emph{header} and \emph{payloads} (\secref{sec:design-extensions}).
\figref{fig:actions} illustrates the action unit microarchitecture.

\begin{figure}[h!]
    \centering
    \includegraphics[width=1.0\linewidth]{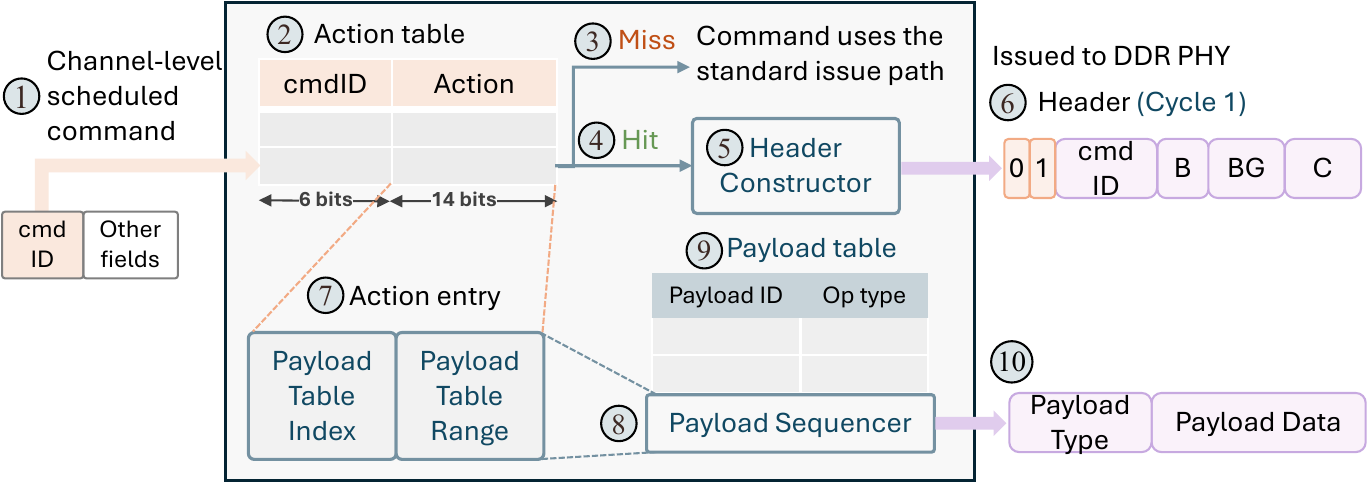}
    \caption{Microarchitecture of the action unit.}
    \label{fig:actions}
\end{figure}

\noindent For every channel-level scheduled command~\cCircled{1}, the action unit uses the \field{cmdID} field as the tag to search the \emph{action table}~\cCircled{2}, a programmable, fully associative structure containing 16 \emph{action entries}.
On an action table miss~\cCircled{3}, the command proceeds through the standard issue path.
On a hit~\cCircled{4}, the action unit performs two steps.
First, the action unit uses the header constructor~\cCircled{5} to populate the header based on the custom command encoding~\cCircled{6} and dispatches it to the DDR PHY.
Second, the action unit constructs payloads and sequences them over the C/A bus using the two attributes of the action entry~\cCircled{7}: (i)~\emph{payload table index} and (ii)~\emph{payload table range}.
These attributes direct the payload sequencer~\cCircled{8} to a contiguous set of entries in the \emph{payload table}~\cCircled{9}, a programmable structure with 128 entries that maps fields of the scheduled command (e.g., row, column, subarray) to the payload identifiers defined by the DRAM vendor (\secref{sec:design-extensions}).
For each selected entry, the payload sequencer extracts the operand from the scheduled command, pairs it with its payload identifier, and dispatches the resulting payload to the DDR PHY in the cycles following the header~\cCircled{10}.

\vspace{2pt}
\noindent\textbf{Example:} PRADA~\cite{shin2024prada,shin2023thesis} uses actions to issue the custom command \texttt{A$^*$}.
On an \texttt{A$^*$} command, the action table returns an action entry.
The action unit constructs the header to specify the \field{cmdID} and the DRAM locality targeted, and subsequently sequences the payloads that identify the subarray and row targeted by the \texttt{A$^*$} command.

\subsection{System Support}
\label{sec:system-support}

To support post-silicon flexibility, \titlename{} leverages system firmware to program the memory controller during the Pre-EFI Initialization (PEI) phase~\cite{uefiplatforminitv19}.
This industry-standard early boot stage occurs before the system memory is initialized, allowing the processor to execute code directly from the motherboard's Serial Peripheral Interface flash (SPI flash)~\cite{intel2023intel}.
\titlename{} programs the memory controller in three steps.
First, the firmware queries the Serial Presence Detect (SPD)~\citeddrstd{} of each installed Dual In-line Memory Module (DIMM) to identify its vendor and module serial number.
Second, the firmware matches the module against a repository of \emph{configurations} signed by the system designer and stored in the SPI flash.
Each configuration contains the specific trigger, update, action, and payload table entries, along with timer update values for that DIMM.
Third, the firmware programs the controller's tables with the identified configuration via Memory-Mapped I/O~\cite{intel32intel}, which provides a standard interface for the CPU to configure the controller's internal logic before the DRAM is online.
In cases where a module's SPD information is unrecognized or a configuration is missing, the controller safely defaults to a transparent JEDEC-compliant mode, which preserves system stability and backward compatibility with standard memory.
As new DIMM modules and security vulnerabilities emerge, a firmware update can deploy new logic without hardware changes.
This follows the established model for deploying microcode updates in modern processors~\cite{intel32intel}.

\subsection{Security}
\label{sec:security}

\titlename{}'s programmable memory controller does \emph{not} expand the system's attack surface beyond that of a dedicated custom memory controller.
Our threat model assumes that the system designer and the DRAM vendor are \emph{trusted} entities.
The attacker attempts to alter the memory controller's programmable tables, for example, to disable a security technique, but \emph{cannot} decrypt secure communication.
Such an attacker \emph{cannot} alter the tables for four reasons.
First, the DIMM configurations (\secref{sec:system-support}) are authored and signed by the system designer, who already provides the platform firmware and is trusted.
Second, the platform's Hardware Root of Trust~\cite{cooper2011bios} maintains the integrity of the DIMM configurations, by ensuring that only cryptographically signed configurations are programmed into the memory controller.
Third, \emph{only} the Pre-EFI Initialization (PEI) system software can program \titlename{}'s programmable tables (\secref{sec:system-support}).
After the PEI phase completes, register lock bits~\cite{intel2019intel} lock the Memory-Mapped I/O registers backing the tables until the next platform reset, so no malicious operating system or user application can modify them.
Fourth, a custom command provides no more control over the DRAM than the dedicated command it replaces, because the DRAM vendor still defines what each command does (\secref{sec:design-extensions}).

\subsection{Hardware Complexity}
\label{sec:overheads}

\vspace{1pt}
\noindent\textbf{DRAM Interface.}
\titlename{} leverages DDR5's existing reserved command encoding space (\secref{sec:design-extensions}) and introduces \emph{no} additional physical pins to the DRAM C/A bus.

\vspace{1pt}
\noindent\textbf{Memory Controller.}
To estimate area, power, and latency overheads, we implement \titlename{} in RTL with support for up to four techniques.
We synthesize the design to GDSII using the OpenROAD~flow~\cite{openroad} and the Nangate 45nm library~\cite{nangate45}, and scale the results to a 10nm node using DeepScale~\cite{satyabrata2021deepscaletool}.
The synthesized design occupies 0.083mm$^2$ and consumes 324mW of static power per DDR5 channel, which is \emph{only} $\sim$0.04\% of the chip area and $\sim$0.67\% of the power budget of a 10nm Intel Xeon processor~\cite{intelemeraldrapids}.
Of the three primitives, triggers add two cycles to the critical path of a memory controller that does not support any techniques.
Updates remain off the critical path for techniques whose minimum timing constraint exceeds the two-cycle metadata read-modify-write latency.
Actions are off the critical path because the custom and standard command issue paths operate in parallel.

\vspace{1pt}
\noindent\textbf{DRAM Logic.}
\titlename{} confines the DRAM-side custom command decoding logic to the DRAM chip center stripe~\cite{keeth2007dram}.
Although individual DRAM techniques may require modifications to the bank peripherals or cell arrays, the custom command extensions require changes only to the center stripe.
The custom command decoder comprises a finite-state machine with one bit of state, a 4-bit payload counter to track up to 16 payloads per custom command (\secref{sec:design-extensions}), and fewer than 24 gates for combinational control logic.
For context, a single 8-bit DDR5 mode register requires eight flip-flops, and the standard~\cite{jedecddr5,jedecddr5c} defines 256 such registers.
By restricting the custom command decoder to the center stripe, \titlename{} does \emph{not} significantly increase DRAM manufacturing complexity beyond what the DRAM techniques themselves require.

\section{Evaluation Methodology}
\label{sec:evaluation-methodology}
We evaluate whether \titlename{} retains the benefits that various DRAM techniques provide with customized implementations.
To do so, we select one representative technique each from four domains of DRAM techniques:
(i)~processing-using-DRAM~(PRADA~\cite{shin2024prada,shin2023thesis}, \secref{sec:processing-using-dram}),
(ii)~lower-overhead~DRAM~maintenance~operations~(MoPAC-C~\cite{vittal2025mopac}, \secref{sec:maintenance}),
(iii)~subarray-level~parallelism~(MASA~\cite{kim2012case}, \secref{sec:parallelism}),
and (iv)~access~latency reduction~(ChargeCache~\cite{hassan2016chargecache}, \secref{sec:memory-latency}).

\vspace{1pt}
\noindent \textbf{Simulator.}
We conduct cycle-level performance and energy simulations using Ramulator~2.0~\citeramulator{}, integrated with DRAMPower~\citedrampower{}.
\tabref{tab:baselineconfigs} presents the baseline system configuration used in our evaluation, unless stated otherwise.

\begin{table}[h!]
    \scriptsize
    \centering
    \caption{Baseline System Configuration}
    \label{tab:baselineconfigs}
    \begin{tabular}{|>{\bfseries}p{1.2cm}|p{6.5cm}|}
    \hline
    \textbf{Processor} & 4.2 GHz, 1- or 4-core, 4-wide issue, 256-entry instruction window \\
    \hline
    \textbf{LLC} & 64B cache-line, 16-way set-associative, 2 MiB/core \\
    \hline
    \multirow{3}{1.2cm}{\textbf{Memory Controller}}
    & 64-entry read/write request queues \\
    & Scheduling policy: FR-FCFS + column cap of 4~\cite{rixner2000memory, zuravleff1997controller, mutlu2007stall} \\
    & Address mapping: RoBaRaCoCh \\
    \hline
    \textbf{DRAM} & DDR5-3200~\cite{jedecddr5,jedecddr5c}, 1 channel, 1 rank, 8 bank groups, 4 banks/bank group, 128 subarrays/bank, 64K rows/bank \\
    \hline
\end{tabular}

\end{table}

We {further} extend Ramulator~2.0 to model:
(i)~\titlename{}-based implementations of the four techniques described in \secref{sec:processing-using-dram}--\secref{sec:memory-latency},
(ii)~custom hardware implementations for these techniques, and
(iii)~\emph{ideal} implementations of these techniques, wherever applicable.
To ensure a fair comparison, we rigorously model the latency overheads of each configuration.
For custom implementations of DRAM techniques, we incorporate latency overheads from their respective memory controller RTL models and optimistically assume 2 cycles for non-standard DRAM commands introduced by these techniques.
For the \titlename{}-based implementations, we incorporate the two cycle memory controller latency overhead derived from our RTL synthesis (\secref{sec:overheads}) and determine their command latencies based on the number of payloads they require (\secref{sec:design-extensions}).

\vspace{1pt}
\noindent \textbf{Workloads.}
We evaluate 57 single-core workloads from five benchmark suites: SPEC CPU2006~\cite{spec2006}, SPEC CPU2017~\cite{spec2017}, TPC~\cite{tpc}, MediaBench~\cite{fritts2009media}, and YCSB~\cite{ycsb}.
We group these single-core workloads into two memory-intensity groups based on their row buffer misses per kilo instructions (RBMPKI): \underline{L}ow ($\text{RBMPKI}{<}2$) and \underline{H}igh ($\text{RBMPKI}{\ge}2$).
We combine randomly selected single-core workloads to construct 60 multi-programmed four-core mixes belonging to five categories, HHHH, HHHL, HHLL, HLLL, and LLLL, with 12 mixes per category.
For PRADA, we curate five microbenchmarks (discussed in more detail in~\secref{sec:processing-using-dram}).

\vspace{1pt}
\noindent \textbf{Metrics.}
We use Instructions Per Cycle (IPC) and weighted speedup~\citeweightedspeedup{}
to evaluate single- and multi-core system performance, respectively.

\section{Evaluation}
\label{sec:evaluation}

In this section, we:
(i)~discuss the four DRAM techniques we evaluate and compare the performance of their \titlename{}-based implementations to their custom implementations (\secref{sec:processing-using-dram}--\secref{sec:memory-latency});
(ii)~show how \titlename{} can compose the performance benefits of multiple techniques by combining ChargeCache and MASA (\secref{sec:putting-together});
(iii)~analyze \titlename{}'s energy (\secref{sec:energy}), latency (\secref{sec:latency}), and area and power (\secref{sec:area-power-tradeoffs}) overheads relative to custom implementations;
(iv)~evaluate \titlename{}'s performance sensitivity to memory controller latency overhead (\secref{sec:sensitivity});
(v)~analyze an alternative \titlename{} implementation that uses Coarse-Grained Reconfigurable Arrays (CGRAs)~\cite{karunaratne2017hycube,wang2019hycube} to realize trigger and update units, trading area and latency for more expressive matching and update logic (\secref{sec:cgra-comparison}).

\subsection{Processing-using-DRAM: PRADA}
\label{sec:processing-using-dram}

Processing-using-DRAM (PuD) is an emerging paradigm that alleviates the data movement bottleneck~\citedatamovement{} between processing elements and main memory by performing massively parallel in-DRAM operations~\citepum{}, including data copy~\cite{seshadri2013rowclone,olgun2022pidram,pidramgithub} and bitwise operations~\cite{seshadri2017ambit,seshadri2016buddy,seshadri2015fast,seshadri2019dram}.
These bitwise operations are performed by simultaneously activating multiple rows in a subarray.
Ambit~\cite{seshadri2017ambit,seshadri2015fast} enables in-DRAM bulk bitwise computation but requires two costly DRAM modifications:
(i)~split row decoders to activate multiple rows simultaneously within a subarray and
(ii)~dual-contact cells to perform NOT operations.

\vspace{1pt}
\noindent\textbf{{PRADA.}}
A recent work, PRADA~\cite{shin2024prada,shin2023thesis}, eliminates the need for both circuit modifications required by Ambit through two mechanisms.
First, Sequential Row Activation (SRA) sequentially activates wordlines to share charge before sensing, obviating the need for a split row decoder.
Second, PRADA introduces two new bitline sense amplifier states that enable NOT operations without dual-contact cells.
PRADA introduces four DRAM commands to support SRA and NOT operations, including \texttt{A$^*$} for activation without sensing and \texttt{N} for NOT.

\vspace{1pt}
\noindent\textbf{{\titlename{}-based PRADA.}}
\titlename{} implements PRADA's four new DRAM commands using its custom command extensions.
Each custom command conveys two payloads: (i)~a subarray address and (ii)~a row address.
We configure \titlename{}'s controller primitives as follows.
{The trigger unit} identifies PuD requests, such as COPY, NOT, AND/OR, NAND/NOR, and XOR/XNOR.
For each request, it sets the \field{cmdID} to the corresponding custom command ID and provides feedback to the scheduler to indicate which command should follow.
{The update unit} maintains timers that enforce the timing constraints of SRA and NOT operations.
{The action unit} issues the custom command sequences required for each PuD request.

\vspace{1pt}
\noindent\textbf{{Results.}}
We compare the performance of the \titlename{}-based and custom implementations of PRADA for in-DRAM computation.
To evaluate how bank-level parallelism affects their performance, we run operations in one~DRAM bank and across 32~DRAM banks.
We evaluate five in-DRAM computation operations: (i)~COPY, (ii)~NOT, (iii)~AND/OR, (iv)~NAND/NOR, and (v)~XOR/XNOR.
For each operation, we run a microbenchmark\footnote{We design a single microbenchmark for the AND and OR operations because they require the same command sequence.
This also applies to NAND/NOR and XOR/XNOR~\cite{shin2024prada,shin2023thesis}.} that repeatedly performs the operation on large input vectors.
We hold the input size per bank constant at 1~MB, resulting in total input sizes of 1~MB and 32~MB for the one DRAM bank and 32 DRAM bank configurations, respectively.
For the CPU baseline, we run equivalent microbenchmarks that perform the same operations on the same input sizes.
\figref{fig:xprada-perf} shows the performance of the \titlename{}-based PRADA implementation (labeled \titlename{}-PRADA) and a custom PRADA implementation (labeled PRADA) for in-DRAM computation operations in 1~bank (left) and across 32~banks (right) normalized to the CPU baseline.

\begin{figure}[h]
    \centering
    \includegraphics[width=1.0\linewidth]{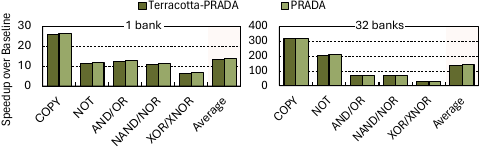}
    \caption{Performance of \titlename{}-PRADA and custom PRADA
    for in-DRAM computation operations in 1 bank (left) and across 32 banks (right).}
    \label{fig:xprada-perf}
\end{figure}

We draw two observations.
First, for in-DRAM computation operations in 1~bank, \titlename{}-PRADA closely matches PRADA's performance.
\titlename{}-PRADA provides an average speedup of 13.36$\times$ over the CPU baseline, compared to 13.79$\times$ for PRADA, retaining 96.9\% of PRADA's average speedup.
We attribute this gap to: (i)~the memory controller latency overhead and (ii)~three-cycle custom command latency (one header and two payloads) compared to custom PRADA's two cycles (\secref{sec:evaluation-methodology}).
Second, for in-DRAM computation operations across 32~banks, \titlename{}-PRADA provides \emph{nearly identical} performance to PRADA.\footnote{For COPY, NOT, and AND/OR microbenchmarks,
\titlename{}-PRADA outperforms PRADA by 0.2\%--1.7\%. We attribute these marginal gains to more favorable memory request scheduling resulting from differences in memory controller and command latencies between \titlename{}-PRADA and PRADA.}
The smaller performance gap indicates that scheduling independent work across banks largely hides the controller and command latency overheads exposed in the 1 bank case.
We conclude that \titlename{}'s latency overheads have a small performance impact on PRADA and that high bank-level parallelism can hide these overheads.

\subsection{Lower-Overhead DRAM Maintenance Operations: MoPAC-C}
\label{sec:maintenance}

RowHammer~\cite{kim2014flipping,mutlu2023retrospectiveflipping,mutlu2019retrospective,kim2020revisiting,mutlu2017rowhammer,cojocar2020we,mutlu2023fundamentally,mutlu2019rowhammer} occurs when a DRAM row is repeatedly activated more times than a pre-defined RowHammer threshold (T\textsubscript{RH}) within a refresh interval, leading to bitflips in physically nearby victim rows~\cite{kim2014flipping}.
A simple way to prevent bitflips is to preventively refresh potential victim rows using refresh commands such as \texttt{REF} or specific preventive refresh commands such as \texttt{RFM}~\cite{jedecddr5}.
To mitigate RowHammer, the April 2024 DDR5 specification~\cite{jedecddr5c} introduced Per-Row Activation Counting (PRAC)~\cite{canpolat2025chronus,canpolat2024understanding}, a mitigation framework that maintains an activation counter for each DRAM row~\cite{kim2014flipping,mutlu2023retrospectiveflipping,mutlu2019retrospective,bennett2021panopticon,bains2016distributed} to accurately determine when to \emph{preventively refresh} victim rows.
PRAC increments each row's activation counter during precharge. When an activation counter reaches T\textsubscript{RH}, PRAC triggers preventive refreshes using \texttt{RFM} commands~\cite{jedecddr5} similar to approaches proposed in prior works~\cite{kim2022mithril,hassan2024smd,devaux2021method,yaglikci2021security,bennett2021panopticon}.
However, updating activation counters during precharge increases the row cycle time (\texttt{tRC}) and precharge latency (\texttt{tRP}), leading to significant performance degradation~\cite{canpolat2024understanding,canpolat2025chronus}.

\vspace{1pt}
\noindent\textbf{{MoPAC-C.}}
To alleviate this performance degradation, MoPAC~\cite{vittal2025mopac} proposes MoPAC-C, a memory controller-side technique that skips PRAC counter updates with a probability $p$.
MoPAC-C preserves security guarantees by carefully selecting $p$ based on the system's RowHammer threshold (T\textsubscript{RH}).
Implementing MoPAC-C requires a non-standard DRAM command, precharge with counter update (\texttt{PREcu}), which the memory controller issues only when MoPAC-C selects a counter update.

\vspace{1pt}
\noindent\textbf{{Terracotta-based MoPAC-C.}}
\titlename{} implements MoPAC-C's \texttt{PREcu} command using its custom command extensions.
The custom \texttt{PREcu} command requires a single payload to convey the subarray address.
\titlename{} tracks pending updates in the metadata table.
We configure \titlename{}'s controller primitives as follows.
{The update unit} sets a metadata table entry with probability $p$ whenever an \texttt{ACT} is scheduled.
{The trigger unit} monitors incoming requests and sets the \field{cmdID} of \texttt{PRE} commands to \texttt{PREcu} if the metadata table entry is set.
The {action unit} issues the custom \texttt{PREcu} command.

\vspace{1pt}
\noindent\textbf{{Results.}}
\figref{fig:xmopac-perf} shows the performance overheads of the \titlename{}-based MoPAC-C implementation, a custom MoPAC-C implementation, and PRAC~\cite{canpolat2025chronus,chronusgithub,canpolat2024understanding} for single-core workloads in subplot~(a) and four-core workload mixes in subplot~(b).
The x-axis shows the workload types with varying memory intensities and the y-axis shows mean performance normalized to a baseline with \emph{no} RowHammer mitigation.
We evaluate all three implementations at three RowHammer thresholds (T\textsubscript{RH}): 250, 500, and 1000.
In each workload type, the patterned bars show the mean performance of \titlename{}-based MoPAC-C (labeled \titlename{}-\{T\textsubscript{RH}\}),
solid bars show the mean performance of custom MoPAC-C (labeled MoPAC-\{T\textsubscript{RH}\}),
the light gray bar shows the mean performance of PRAC, and error bars show the standard error of the mean across all the workloads of the workload type.
We show a single bar for PRAC because its performance does \emph{not} vary across the evaluated thresholds~\cite{canpolat2025chronus}.
\emph{GMean} shows the geometric mean of normalized performance for each implementation across the workload types.

\begin{figure}[h!]
    \centering
    \includegraphics[width=1.0\linewidth]{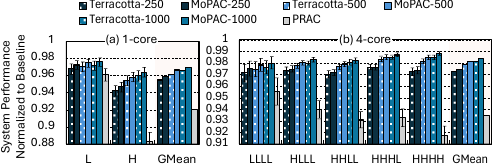}
    \caption{Performance impact of \titlename{} MoPAC-C, custom MoPAC-C, and PRAC for three RowHammer thresholds: 250, 500, and 1000, on single-core and four-core workload mixes.}
    \label{fig:xmopac-perf}
\end{figure}

We draw three key observations.
First, for single-core workloads and four-core workload mixes, \titlename{}-based MoPAC-C closely matches the performance of MoPAC-C across all three RowHammer thresholds.
\titlename{}-based MoPAC-C provides geometric mean performance within 0.3\% (0.5\%) of custom MoPAC-C for four-core workload mixes (single-core workloads).
Second, \titlename{}-based MoPAC-C recovers a significant fraction of performance lost to PRAC across all T\textsubscript{RH} values.
At the lowest evaluated RowHammer threshold, 250, \titlename{}-based MoPAC-C provides normalized geometric mean performance of 0.97 (0.95) compared to PRAC's 0.93 (0.92) for four-core workload mixes (single-core workloads).
This corresponds to a 4.1\% (3.7\%) performance improvement over PRAC for four-core workload mixes (single-core workloads).
Third, \titlename{}-based MoPAC-C retains a large fraction of the performance benefits of custom MoPAC-C over PRAC, retaining 96.0\% (88.2\%) of the custom implementation's improvement for four-core workload mixes (single-core workloads) at a RowHammer threshold of 250.
We conclude that \titlename{} can effectively enable the benefits of techniques that perform new DRAM maintenance operations.

\subsection{DRAM Parallelism: MASA}
\label{sec:parallelism}

To serve multiple memory requests in parallel, modern DRAM systems employ multiple banks that can be accessed independently.
However, memory requests to different rows in the same bank incur \emph{bank conflicts}~\cite{kim2012case,rau1991pseudo}, requiring the memory controller to close the open row before activating another.
Such bank conflicts can lead to a serialization of memory requests and row buffer thrashing, degrading performance~\cite{kim2012case}.
While increasing the number of DRAM banks can reduce bank conflicts, this comes at a high cost~\cite{kim2012case}.

\vspace{1pt}
\noindent\textbf{{MASA.}}
To mitigate the performance impact of {bank conflicts},
MASA (Multitude of Activated SubArrays)~\cite{kim2012case}
exposes DRAM subarrays to the memory controller and allows multiple subarrays within a bank to remain active simultaneously, improving DRAM parallelism.
MASA's memory controller uses a new subarray select command (\texttt{SA\_SEL}) to designate the subarray that will serve the next read (\texttt{RD}) or write (\texttt{WR}) command.
Since MASA exposes subarrays to the memory controller, it also provides the memory controller with a selective precharge (\texttt{PRE\_SA}) command to precharge an individual subarray without affecting other subarrays in the same bank.

\vspace{1pt}
\noindent\textbf{{\titlename{}-based MASA.}}
\titlename{} implements MASA's \texttt{SA\_SEL} and \texttt{PRE\_SA} commands using its custom command extensions.
Each custom command requires a single payload to convey the subarray address.
\titlename{} tracks the currently selected subarray of each bank by utilizing the metadata table.
We configure \titlename{}'s controller primitives as follows.
{The trigger unit} uses the tracked selected subarray to fire the two custom commands.
It fires a subarray select when a scheduled request targets an active but non-selected subarray,
and it fires a selective precharge when a scheduled request targets a different row in an already active subarray.
{The update unit} updates the selected subarray in the metadata table on every subarray select or selective precharge.
{The action unit} issues the custom \texttt{SA\_SEL} and \texttt{PRE\_SA} commands.

\vspace{1pt}
\noindent\textbf{{Results.}}
\figref{fig:xmasa-perf} shows the performance of the \titlename{}-based MASA implementation, a custom MASA implementation, and an ideal DRAM system where every subarray is modeled as a DRAM bank (``Ideal''), for single-core workloads in subplot~(a) and four-core workload mixes in subplot~(b).
The x-axis shows the workload types with varying memory intensities and the y-axis shows speedup normalized to a baseline without subarray-level parallelism.
In each workload type, the three bars from left to right show the mean performance of \titlename{}-based MASA (labeled \titlename{}-MASA), custom MASA (labeled MASA), and the ideal DRAM system (labeled ``Ideal''),
and error bars show the standard error of the mean across all the workloads of the workload type.
\emph{GMean} shows the geometric mean performance of each implementation across the workloads of all types.

\begin{figure}[h!]
    \centering
    \includegraphics[width=1.0\linewidth]{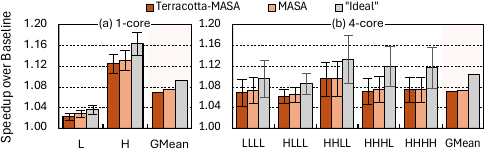}
    \caption{Performance of \titlename{}-MASA, custom MASA, and an ideal DRAM system on single-core workloads and four-core workload mixes.}
    \label{fig:xmasa-perf}
\end{figure}

We make the following two observations.
First, for both single-core workloads and four-core workload mixes, \titlename{}-MASA closely matches the speedup of MASA over the baseline across all workload types.
\titlename{}-MASA's geometric mean performance is only 0.22\% (0.51\%) lower than MASA for four-core workload mixes (single-core workloads), retaining 96.72\% (92.66\%) of MASA's performance improvement over the baseline.
Second, \titlename{}-MASA captures a similar fraction of the ``Ideal'' DRAM system's performance improvement as MASA.
For four-core workload mixes (single-core workloads), \titlename{}-MASA provides 68.3\% (74.6\%) of ``Ideal'''s geometric mean performance improvement over the baseline, compared to 70.7\% (80.5\%) for custom MASA.
We conclude that \titlename{} can effectively enable the benefits of techniques that improve DRAM parallelism.

\subsection{Reducing Memory Latency: ChargeCache}
\label{sec:memory-latency}

The memory controller enforces a fixed set of DRAM timing constraints (e.g., \texttt{tRCD}) that the manufacturer sets conservatively to guarantee correct operation for the slowest cell under worst-case conditions~\cite{chang2016understanding,lee2017design,kim2018solar,lee2015adaptive}.
Since DRAM is heavily optimized for cost-per-bit, these timing constraints, and thus DRAM access latencies, have remained largely stagnant across DRAM generations~\cite{patel2022case,patel2024rethinking,son2013reducing,chang2017thesis,lee2013tiered,lee2016reducing,lee2015adaptive,lee2017design}.

\vspace{1pt}
\noindent\textbf{{ChargeCache.}}
ChargeCache~\cite{hassan2016chargecache} reduces DRAM access latency by leveraging two observations.
First, the cells of a row that was recently precharged remain in a ``highly charged'' state.
Second, highly charged cells can be accessed faster than cells at a lower charge level, so an activation to such a row can use reduced timing constraints.
ChargeCache's memory controller tracks the addresses of recently precharged rows in a small controller-side table and serves subsequent activations to those rows with reduced timing constraints.

\vspace{1pt}
\noindent\textbf{{\titlename{}-based ChargeCache.}}
\titlename{} uses its metadata table to track recently precharged rows.
We configure \titlename{}'s controller primitives as follows.
{The trigger unit} performs a metadata lookup when an activation (\texttt{ACT}) is scheduled.
On a metadata hit, it marks the request's technique ID (\field{techID}) field to signal a faster access.
{The update unit} performs two tasks.
Upon a precharge (\texttt{PRE}), it inserts the open row's address into the metadata table to track its highly charged state.
Upon an \texttt{ACT} to a tracked row, it loads the relevant timers with the reduced timing constraints.
{The action unit} is not used because ChargeCache requires no custom commands.

\vspace{1pt}
\noindent\textbf{{Results.}}
\figref{fig:xcc-perf} shows the performance of the \titlename{}-based ChargeCache implementation, a custom ChargeCache implementation, and an ideal DRAM system where every activation uses reduced timing constraints, for single-core workloads in subplot~(a) and four-core workload mixes in subplot~(b).
The x-axis shows the workload types with varying memory intensities and the y-axis shows speedup normalized to the baseline.
In each workload type, the three bars from left to right show the mean performance of \titlename{}-based ChargeCache (labeled \titlename{}-ChargeCache), custom ChargeCache (labeled ChargeCache), and the ideal DRAM system (labeled ``Low-Latency DRAM''),
and error bars show the standard error of the mean across all the workloads of the workload type.
\emph{GMean} shows the geometric mean performance of each implementation across the workloads of all types.

\begin{figure}[h!]
    \centering
    \includegraphics[width=1.0\linewidth]{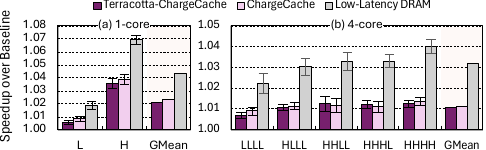}
    \caption{Performance of \titlename{}-ChargeCache, custom ChargeCache, and a low-latency DRAM system on single-core workloads and four-core workload mixes.}
    \label{fig:xcc-perf}
\end{figure}

We observe that although a custom ChargeCache provides modest performance improvements,
\titlename{}-ChargeCache retains 96.4\% (87.7\%) of these benefits,
providing a geometric mean speedup of 1.09\% (2.04\%) over the baseline for four-core workload mixes (single-core workloads).
We conclude that \titlename{} maintains most of the performance benefits of memory access latency reduction techniques such as ChargeCache despite its memory controller latency overheads.

\subsection{Composing ChargeCache and MASA}
\label{sec:putting-together}

With \titlename{}, a system initially deployed with one technique can later adopt another technique supported by its existing hardware via a firmware update.
We leverage \titlename{} to compose ChargeCache (\secref{sec:memory-latency}) and MASA (\secref{sec:parallelism}) into \emph{ChargeCacheMASA} and show that \titlename{} can retain the benefits of their composition.
To configure {ChargeCacheMASA}, we configure the two techniques and also set the trigger unit's C/A bus utilization filter threshold and trigger priorities (\secref{sec:design-triggers}).

\vspace{1pt}
\noindent\textbf{{Results.}}
\figref{fig:xccmasa-perf} shows the performance of \titlename{}-based and custom implementations of ChargeCache, MASA, and ChargeCacheMASA,\footnote{Among the configurations evaluated, we find the best performance by setting the C/A bus utilization filter threshold to 40\% utilization (\secref{sec:design-triggers}) and prioritizing MASA over ChargeCache when both remain eligible for a request (\secref{sec:design-triggers}).} for single-core workloads in subplot~(a) and four-core workload mixes in subplot~(b).
The x-axis shows the workload types with varying memory intensities and the y-axis shows speedup normalized to the baseline.
In each workload type, the three pairs of bars from left to right show ChargeCache, MASA, and ChargeCacheMASA.
Within each pair, the left bar shows the mean performance of the \titlename{}-based implementation (labeled \titlename{}-\{technique\}) and the right bar shows the custom implementation (labeled \{technique\}),
and error bars show the standard error of the mean across all the workloads of the workload type.
\emph{GMean} shows the geometric mean performance of each implementation across the workloads of all types.

\begin{figure}[h!]
    \centering
    \includegraphics[width=1.0\linewidth]{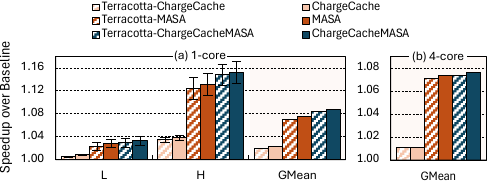}
    \caption{Performance of \titlename{}-based and custom implementations of ChargeCache, MASA, and their combination (ChargeCacheMASA) on single-core workloads and four-core workload mixes with varying memory intensities.}
    \label{fig:xccmasa-perf}
\end{figure}

We draw two key observations.
First, the \titlename{} implementation of ChargeCacheMASA provides performance benefits over the \titlename{} implementations of both individual techniques.
For single-core workloads, it provides speedups of 6.4\% and 1.4\% over \titlename{}-ChargeCache and \titlename{}-MASA, respectively.
For four-core workload mixes, the corresponding speedups are 6.1\% and 0.2\%.
The benefits of combining ChargeCache and MASA are \emph{sub-additive}, especially for four-core workload mixes, because MASA keeps more rows activated, reducing the opportunities for ChargeCache to accelerate accesses to recently precharged rows.
Second, \titlename{}-ChargeCacheMASA closely matches the performance of ChargeCacheMASA, retaining 95.9\% (95.8\%) of its performance improvement over the baseline for four-core workload mixes (single-core workloads).
We conclude that \titlename{} preserves the benefits of combining techniques and enables a system to improve performance by adding additional techniques through firmware updates.

\subsection{Overheads of \titlename{}}

\subsubsection{Energy}
\label{sec:energy}

We evaluate the DRAM energy consumption of the four techniques (\secref{sec:processing-using-dram}--\secref{sec:memory-latency}) and their composition (\secref{sec:putting-together}).
Here, we report \emph{only} the DRAM energy overhead introduced by \titlename{} relative to custom implementations.
\figref{fig:energy-overheads} shows this overhead for each technique and their composition.

\begin{figure}[h!]
    \centering
    \includegraphics[width=1.0\linewidth]{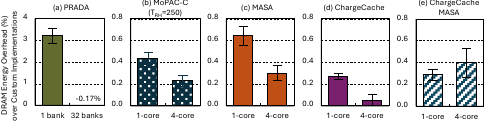}
    \caption{DRAM energy overhead (\%) of \titlename{} implementations over custom implementations. Note the different y-axis scales across subplots.}
    \label{fig:energy-overheads}
\end{figure}

We observe that \titlename{}-based PRADA consumes 3.2\% more DRAM energy than custom PRADA when performing in-DRAM computation operations in 1~bank, but reduces average DRAM energy consumption by 0.17\% when operating across 32~banks.
We attribute this reduction to more favorable memory request scheduling arising from differences in controller and command latencies, which slightly shorten execution time.
For the remaining techniques and their composition, the average DRAM energy overhead remains below 0.64\% (0.40\%) for single-core workloads (four-core workload mixes).

\subsubsection{Memory Access Latency}
\label{sec:latency}

We evaluate \titlename{}'s impact on memory access latency
by comparing the 99th-percentile memory access latency of \titlename{} implementations with custom implementations.
We evaluate MoPAC-C (at T\textsubscript{RH}=250), MASA, and ChargeCache using four-core workload mixes, and PRADA using in-DRAM computation operations across 32 banks.
We observe that \titlename{} increases the 99th-percentile memory access latency by \emph{only} 0.17\%.

\subsubsection{Area and Power}
\label{sec:area-power-tradeoffs}

Our RTL synthesis results, obtained using the methodology in \secref{sec:overheads},
show that \titlename{} incurs modest area and power overheads over a design with custom controller support for the evaluated DRAM techniques.
These additional area and power amount to \emph{only} 0.03\% of the die area and 0.56\% of the thermal design power (TDP) of a high-end Intel Xeon processor~\cite{intelemeraldrapids}.
These overheads constitute a one-time implementation cost that can be amortized across the techniques deployed over the processor's lifetime.

\subsubsection{Sensitivity to Controller Latency Overhead}
\label{sec:sensitivity}

\titlename{}'s trigger unit adds a controller latency overhead of two cycles (\secref{sec:overheads}).
To examine how a higher controller latency affects performance, we sweep this overhead from two cycles to 10 cycles in increments of two cycles for the four techniques (\secref{sec:processing-using-dram}--\secref{sec:memory-latency}) and their composition (\secref{sec:putting-together}).
\figref{fig:latency-sensitivity} shows their performance relative to their respective baselines.
We evaluate PRADA using in-DRAM computation operations across 32~banks (subplot~(a)) and the remaining techniques and their composition using four-core workload mixes (subplots~(b)--(e)).

\begin{figure}[h!]
    \centering
    \includegraphics[width=\linewidth]{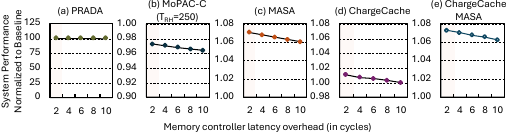}
    \caption{Performance sensitivity of Terracotta implementations to memory controller latency overhead. Note the different y-axis scales across subplots.}
    \label{fig:latency-sensitivity}
\end{figure}

We draw two observations.
First, PRADA's performance remains unchanged with increase in memory controller latency overhead.
We attribute this to the high bank-level parallelism exposed by our microbenchmarks, which perform independent in-DRAM computation operations across 32~banks.
Second, the other techniques and their composition show a marginal performance decrease across the evaluated range of memory controller latency overhead.
We conclude that the performance of \titlename{}-based implementations is robust to higher controller latency overheads.

\subsection{\titlename{} with Reconfigurable Logic}
\label{sec:cgra-comparison}

\secref{sec:implementation} describes a concrete implementation of \titlename{} using programmable tables.
In this section, we evaluate the performance implications of an alternative \titlename{} implementation that uses reconfigurable logic.
Specifically, we implement \titlename{}'s trigger and update units using 6$\times$6 HyCUBE-like Coarse-Grained Reconfigurable Arrays (CGRAs)~\cite{karunaratne2017hycube,wang2019hycube}.
We retain programmable tables for the action primitive because issuing custom DRAM commands does \emph{not} require additional flexibility.
We map the trigger and update logic of each technique onto the CGRA using Morpher~\cite{morphergithub,juneja2025building}, a CGRA compiler, to determine the resulting trigger and update latencies, and incorporate them into Ramulator~2.0~\citeramulator{}.

\begin{figure}[!b]
    \centering
    \includegraphics[width=1.0\linewidth]{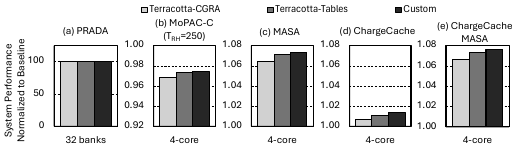}
    \caption{Performance of \titlename{}-CGRA, \titlename{}-Tables, and custom implementations for four techniques and their composition. Note the different y-axis scales across subplots.}
    \label{fig:cgra}
\end{figure}

\vspace{1pt}
\noindent\textbf{{Results.}}
\figref{fig:cgra} shows the performance of the CGRA-based \titlename{} implementation (labeled \titlename{}-CGRA), the programmable table-based \titlename{} implementation (labeled \titlename{}-Tables), and a custom implementation (labeled Custom).
We evaluate PRADA using in-DRAM computation operations across 32~banks (subplot~(a)) and the remaining techniques and their composition using four-core workload mixes (subplots~(b)--(e)).
The y-axes show their speedup normalized to their respective baselines.
In each subplot, the three bars from left to right show the geometric mean performance of \titlename{}-CGRA, \titlename{}-Tables, and Custom.

We observe that \titlename{}-CGRA provides similar performance to custom implementations, with little or no performance loss relative to \titlename{}-Tables.
For PRADA, \titlename{}-CGRA preserves nearly all the performance benefits of the custom implementation.
For the remaining techniques and their composition, its performance is only 0.37--0.66\% lower than \titlename{}-Tables.
We conclude that reconfigurable logic enables support for richer control flow while largely preserving the performance benefits of custom and programmable table-based \titlename{} implementations.

\section{Discussion}
\label{sec:discussion}

\subsection{Extending \titlename{}}
\label{sec:extend}

We derive \titlename{}'s primitives from an analysis of \numanalyzedtechniquescited{} techniques across \numanalyzedcategories{} categories.
However, these primitives may \emph{not} capture the requirements of all possible DRAM techniques.
We discuss cases that may require extending \titlename{} and, where possible, outline potential extensions.
We leave the design and evaluation of these extensions to future work.

\vspace{1pt}
\noindent\textbf{Multi-bank coordination.}
\titlename{} partitions its trigger logic across banks (\secref{sec:design}) and does \emph{not} support coordination across them.
Some Processing-near-Memory (PnM) techniques~\citepnm{}, which add logic \emph{near} memory arrays to alleviate the data movement bottleneck~\citedatamovement{}, require such coordination.
These techniques use a single command to perform the same operation across multiple banks~\cite{kwon2022system,he2020newton,lee20221ynm,yoon2025splitsync} or all banks~\cite{kwon202120nm,lee2021hardware,kim2025pimba}.
At the interface level, \titlename{}'s custom command extensions can already encode operations on predefined groups of banks, including all banks, by allowing the DRAM-side command decoder to interpret the locality fields based on the command's operation type.
Targeting an arbitrary set of banks may require extending the locality primitive.
At the controller level, one potential extension is to introduce channel- or bank group-level triggers to coordinate the firing of techniques across banks.
Such a controller extension may also benefit memory scheduling~\cite{moscibroda2008distributed,kim2010atlas,mutlu2009parallelism} by coordinating memory requests across bank groups and banks to improve system performance and fairness.

\vspace{1pt}
\noindent\textbf{DRAM feedback.}
\titlename{}'s triggers, updates, and actions operate on the scheduled command and stored metadata, but do \emph{not} directly observe the outcomes of DRAM operations.
Techniques that use these outcomes to adapt DRAM operation therefore require additional support.
Examples include techniques that use online profiling for retention-aware refresh~\cite{liu2012raidr,qureshi2015avatar,patel2017reaper,khan2016parbor} and access-time reduction~\cite{lee2015adaptive,kim2018solar,lee2017design}.
Such feedback could also enable online profiling of read disturbance characteristics, including RowHammer thresholds,
their variation across memory locations~\cite{yaglikci2024spatial} and over time~\cite{olgun2025variable},
and their dependence on temperature and row open time~\cite{orosa2021deeper,luo2023rowpress,nam2024dramscope,nam2023xray,luo2025revisiting}.
These profiles could enable future mitigations that adapt their preventive actions to the variation in read disturbance characteristics across chips, time, physical locations, and data patterns~\cite{olgun2026discord,olgun2026qrd}.
One potential extension is a feedback path that records profiling results in \titlename{}'s state, allowing triggers and updates to use these results to adapt subsequent DRAM operations.

\subsection{Limitations of \titlename{}}
\label{sec:limitations}

We identify \titlename{}'s limitations under three categories.

\vspace{1pt}
\noindent\textbf{Technique-specific DRAM modifications.}
Many DRAM techniques require modifications to the DRAM chip itself, in addition to the interface and controller support that \titlename{} provides (\secref{sec:design}).
These modifications are specific to each technique.
Supporting all of them in a single DRAM architecture could add circuitry that individual products might \emph{not} need.
\titlename{} leaves these modifications to DRAM vendors to allow each vendor to implement the techniques they see fit and expose them through custom command extensions.
This allows vendors to introduce technique-specific DRAM functionality while reducing the need for interface and memory controller modifications.

\vspace{1pt}
\noindent\textbf{Technique-specific processor modifications.}
Some techniques~\cite{hashemi2016accelerating,khadem2025dx100,ahn2015pim,hsieh2016transparent,boroumand2019conda,boroumand2017lazypim,vittal2026bard,hashemi20216continuous} require changes to the core or the memory hierarchy to offload work to the DRAM system.
These changes remain the system designer's responsibility and fall outside the flexibility that \titlename{} provides, although \titlename{} can still support the parts of these techniques that need DRAM interface and controller support.

\vspace{1pt}
\noindent\textbf{Implementation constraints.}
A given \titlename{} implementation provides a fixed amount of metadata storage and a fixed number of timers (\secref{sec:design}).
A technique's updates remain off the critical path only if its minimum timing constraint exceeds the metadata table read-modify-write latency (\secref{sec:overheads}).
Techniques that need more resources, or whose timing constraints are tighter than this latency, may require a controller redesign.

\section{Related Work}

To our knowledge, \titlename{} is the first work that reduces the repeated interface and controller modifications needed to enable the adoption of DRAM techniques~\citetechniques{}.

\vspace{1pt}
Self-Managing DRAM (SMD)~\cite{hassan2024smd,hassan2022self} and \titlename{} have different goals.
SMD aims to transfer the responsibility for DRAM maintenance from the memory controller to the DRAM chip, enabling the DRAM chip to autonomously maintain itself.
Its purpose is \emph{not} to provide support for new DRAM techniques that require controller-side changes.
In contrast, \titlename{} aims to enable such DRAM techniques via a flexible DRAM interface and a programmable memory controller by reducing the need for repeated interface and controller modifications.
As such, these frameworks are complementary and can be combined to enable more powerful innovations in the DRAM system~\cite{patel2022case,patel2024rethinking,taneja2026from,saroiu2026from,frigo2020trrespass,jattke2022blacksmith,kim2020revisiting} to mitigate the limitations of modern DRAM-based systems.

\vspace{1pt}
PARDIS~\cite{bojnordi2012pardis} provides programmable control over address resolution, scheduling policies, and power management to improve system performance and reduce energy consumption.
However, PARDIS \emph{cannot} implement techniques that introduce new DRAM commands or maintain metadata in the memory controller.
In contrast, \titlename{} provides a flexible framework to enable techniques and commands beyond the standard, making the two approaches complementary.

\vspace{1pt}
FPGA-based DRAM infrastructures like DRAM Bender~\cite{safari-drambender, olgun2023drambender, olgun2026modern} and SoftMC~\cite{softmc-safarigithub, hassan2017softmc} are instrumental
to understanding DRAM robustness issues~\cite{kim2014flipping,luo2023rowpress,yuksel2025columndisturb,yuksel2025pudhammer,luo2026dejavu,olgun2025variable,tugrul2025understanding,yaglikci2024spatial,kim2020revisiting,tokuda2026pudghost,patel2020beer,patel2021harp,patel2017reaper,qureshi2015avatar},
understanding DRAM latency~\cite{chang2016understanding, chang2017understanding, lee2017design,lee2015adaptive,chang2017thesis,lee2016reducing} and retention~\cite{liu2013experimental,khan2016parbor,patel2017reaper,qureshi2015avatar,khan2014efficacy} characteristics,
uncovering new DRAM functionality~\cite{yuksel2025indram,yuksel2024simultaneous,yuksel2024functionally,yaglikci2022hira,orosa2021codic,olgun2021quac,kim2019drange,garzon2026cadm,kubo2025pudtune,gao2022fracdram,gao2019computedram,tokuda2026clutch,baser2026indram,kim2018the},
and gaining insight into DRAM's internal structure~\cite{chang2017understanding,chang2016understanding,nam2024dramscope,nam2023xray,patel2020beer,patel2021harp,patel2019understanding,khan2016parbor}.
PiDRAM~\cite{olgun2022pidram, pidramgithub} and EasyDRAM~\cite{canpolat2025easydram, easydramgithub} provide end-to-end system integration for some novel DRAM functionalities.
These infrastructures expose functionality through existing DRAM commands.
In contrast, \titlename{} enables the deployment of diverse DRAM techniques through custom command extensions that let DRAM vendors define new commands within a standardized interface extension and a programmable memory controller that provides technique-specific support post-silicon fabrication.

\section{Conclusion}
\label{sec:conclusion}

We propose {\titlename{}}, a new, flexible framework that reduces the repeated DRAM interface and memory controller modifications needed to adopt diverse DRAM techniques
by capturing their similarities in DRAM commands and controller workflows into a set of configurable primitives.
\titlename{} consists of two components:
(i)~\emph{custom command extensions} that enable DRAM vendors to define new commands within a single, standardized interface extension, and
(ii)~a \emph{programmable memory controller}
that system designers can easily and flexibly program to support new DRAM techniques post-silicon fabrication.
We use \titlename{} to implement four DRAM techniques across the domains of processing-using-DRAM, low-cost DRAM maintenance, subarray-level parallelism, and latency reduction.
We show that \titlename{} provides almost all of the performance benefits of their respective custom implementations, at low DRAM energy overheads and controller area and power overheads.
We hope and expect that \titlename{} will inspire academia and industry to develop innovative DRAM techniques and help overcome stagnation in their adoption, enabling large improvements in DRAM-based systems.

\section*{Acknowledgments}

We thank the anonymous reviewers of MICRO SRC 2025, ISCA 2026, and MICRO 2026 for their feedback.
We thank members of the SAFARI Research Group (especially Umut Ba\c{s}er and Vlad-Petru Nitu) for their constructive feedback and for providing a stimulating, intellectual, and scientific environment.
We acknowledge the generous gift funding provided by our industrial partners (especially Google, Huawei, Intel, Microsoft),
which has been instrumental in enabling the research we have been conducting on Processing-in-Memory and read disturbance in DRAM, in particular,
and memory systems in general~\cite{mutlu2019processing,mutlu2022modern,mutlu2025modernprimerprocessingmemory,mutlu2013memory, mutlu2014research, mutlu2017rowhammer, mutlu2025memory,mutlu2019rowhammer,
mutlu2023fundamentally,mutlu2020intelligentdate,mutlu2019enabling,
mutlu2019retrospective,mutlu2023retrospectiveflipping,mutlu2023retrospectiveexperimental,mutlu2023retrospectiveraidr,ahn2023retrospective,ipek2023retrospective,
yuksel2026memory,ghose2019processing,olgun2026modern,mutlu2024memory,bostanci2026extended,luo2026ramulator2.1,yuksel2025columndisturb,yuksel2025pudhammer,luo2026dejavu,luo2023rowpress,kim2014flipping,
kakolyris2026columnkeeper,bera2021pythia,bera2022hermes,kanellopoulos2026revelator, oliveira2022accelerating, bera2026machine}.
This work was in part supported by the Google Security and Privacy Research Award and the Microsoft Swiss Joint Research Center.

\clearpage
\balance
\bibliographystyle{IEEEtran}
\bibliography{refs}

\clearpage
\nobalance
\appendix
\section{Modifications Required by the Analyzed DRAM Techniques}
\label{sec:appendix-modifications}

\begin{table}[!b]
\centering
\footnotesize
\setlength{\tabcolsep}{4pt}
\renewcommand{\arraystretch}{1.05}
\caption{Modifications required by each of the \numanalyzedtechniquescited{} analyzed DRAM techniques. MC = memory controller, IF = DRAM interface, Chip = DRAM chip.}
\label{tab:appendix-modifications}
\begin{tabularx}{\columnwidth}{|X||c|c|c|}
\hline
\textbf{Technique} & \textbf{MC} & \textbf{IF} & \textbf{Chip} \\
\hline\hline
\multicolumn{4}{|l|}{\textit{Processing-using-DRAM}} \\
\hline
RowClone~\cite{seshadri2013rowclone}        & \cmark & \cmark & \cmark \\
Ambit~\cite{seshadri2017ambit}              & \cmark & \xmark & \cmark \\
ComputeDRAM~\cite{gao2019computedram}       & \cmark & \xmark & \xmark \\
FracDRAM~\cite{gao2022fracdram}             & \cmark & \xmark & \xmark \\
PRADA~\cite{shin2024prada}                  & \cmark & \cmark & \cmark \\
\hline
\multicolumn{4}{|l|}{\textit{Probabilistic controller-side read disturbance mitigation}} \\
\hline
PARA~\cite{kim2014flipping}                 & \cmark & \xmark & \xmark \\
MoPAC-C~\cite{vittal2025mopac}                & \cmark & \cmark & \xmark \\
\hline
\multicolumn{4}{|l|}{\textit{Subarray-level parallelism}} \\
\hline
MASA~\cite{kim2012case}                     & \cmark & \cmark & \cmark \\
SARP~\cite{chang2014improving}              & \cmark & \xmark & \cmark \\
HiRA~\cite{yaglikci2022hira}                & \cmark & \xmark & \xmark \\
\hline
\multicolumn{4}{|l|}{\textit{Access latency reduction}} \\
\hline
TL-DRAM~\cite{lee2013tiered}                & \cmark & \xmark & \cmark \\
NUAT~\cite{shin2014nuat}                    & \cmark & \xmark & \xmark \\
DAS-DRAM~\cite{lu2015improving}             & \cmark & \xmark & \cmark \\
ChargeCache~\cite{hassan2016chargecache}    & \cmark & \xmark & \xmark \\
LISA~\cite{chang2016low}                    & \cmark & \cmark & \cmark \\
CROW~\cite{hassan2019crow}                  & \cmark & \cmark & \cmark \\
FIGARO~\cite{wang2020figaro}                & \cmark & \cmark & \cmark \\
FASA-DRAM~\cite{du2024fasa}                 & \cmark & \cmark & \cmark \\
\hline
\multicolumn{4}{|l|}{\textit{Deterministic controller-side read disturbance mitigation}} \\
\hline
Counter-based~\cite{kim2014flipping}        & \cmark & \xmark & \xmark \\
CRA~\cite{kim2014architectural}             & \cmark & \xmark & \xmark \\
CAT~\cite{seyedzadeh2018cbt}                & \cmark & \xmark & \xmark \\
Graphene~\cite{park2020graphene}            & \cmark & \cmark & \xmark \\
Mithril~\cite{kim2022mithril}               & \cmark & \xmark & \cmark \\
AQUA~\cite{saxena2022aqua}                  & \cmark & \xmark & \xmark \\
RRS~\cite{saileshwar2022randomized}         & \cmark & \xmark & \xmark \\
ABACuS~\cite{olgun2024abacus}               & \cmark & \xmark & \xmark \\
CoMeT~\cite{bostanci2024comet}              & \cmark & \xmark & \xmark \\
\hline
\multicolumn{4}{|l|}{\textit{In-DRAM true random number generation}} \\
\hline
D-RaNGe~\cite{kim2019drange}                & \cmark & \xmark & \xmark \\
QUAC-TRNG~\cite{olgun2021quac}              & \cmark & \xmark & \xmark \\
\hline
\multicolumn{4}{|l|}{\textit{Fine-grained activation}} \\
\hline
Half-DRAM~\cite{zhang2014half}              & \cmark & \xmark & \cmark \\
FGDRAM~\cite{oconnor2017fine}               & \cmark & \cmark & \cmark \\
Sectored DRAM~\cite{olgun2024sectored}      & \cmark & \cmark & \cmark \\
\hline\hline
\textbf{Total (of \numanalyzedtechniquescited{})} & \textbf{32} & \textbf{11} & \textbf{15} \\
\hline
\end{tabularx}
\end{table}
This appendix discusses the DRAM system modifications required by the \numanalyzedtechniquescited{} techniques in \tabref{tab:techniques}.
\tabref{tab:appendix-modifications} classifies these modifications into three categories: memory controller (MC), DRAM interface (IF), or DRAM chip (Chip) modifications. We define these categories below.

\vspace{1pt}
\noindent\textbf{Memory controller modification.}
A technique requires a memory controller modification if a conventional memory controller that follows the DRAM standard cannot support it.
Such modifications typically include adding memory controller state (e.g., activation counters or timers), conditions to trigger techniques, or rules for updating controller state.
The controller modifications may also be required to issue new DRAM commands or use existing DRAM commands with timing constraints that the standard does not specify.

\vspace{1pt}
\noindent\textbf{DRAM interface modification.}
A technique requires a DRAM interface modification if it introduces a DRAM command that the DRAM standard does not define or assigns a new functional meaning to bits that the standard reserves or leaves unused.
This definition excludes techniques that issue only standard DRAM commands, even if they issue them in sequences or with timing constraints that the standard does not specify, or if a modified DRAM chip responds to them differently.

\vspace{1pt}
\noindent\textbf{DRAM chip modification.}
A technique requires a DRAM chip modification if it proposes to change the DRAM cell array, its peripheral circuits, or both.
A technique that works on unmodified, commercially available DRAM chips does \emph{not} require a DRAM chip modification.

\section{Artifact Appendix}

\newtcolorbox{cmdbox}{colback=black!2,colframe=black!25,boxrule=0.3pt,
  arc=1pt,left=4pt,right=4pt,top=2pt,bottom=2pt,
  before skip=4pt,after skip=4pt,fontupper=\ttfamily\footnotesize}

\newcommand{\artifactresult}[2]{%
  \par\addvspace{3pt}\noindent{{(#1) #2.}}\enspace\ignorespaces}

\subsection{Abstract}
Our artifact provides the source code, scripts, and data required to reproduce the results of \titlename{}.

\subsubsection{Source code, scripts, and data}
{We provide (i)~a Ramulator~2.0-based cycle-level simulator integrated with DRAMPower and extended to model custom and \titlename{} implementations of PRADA~\cite{shin2024prada,shin2023thesis}, MoPAC-C~\cite{vittal2025mopac}, MASA~\cite{kim2012case}, ChargeCache~\cite{hassan2016chargecache}, and the composition of ChargeCache and MASA;
(ii)~RTL sources for evaluating controller area and power;
(iii)~workload memory traces; and
(iv)~Bash and Python scripts to build the tools, run experiments, parse results, and generate plots.}

\subsubsection{Reproducible results}
{The artifact supports reproduction of the performance and DRAM energy results (\figref{fig:xprada-perf}--\figref{fig:cgra}), memory access latency results (\secref{sec:latency}), and controller area and power results (\secref{sec:area-power-tradeoffs}).}

\subsection{Artifact check-list (meta-information)}
{\small
\begin{itemize}\itemsep2pt
  \item \textbf{Program:} C++ and RTL programs; Python~3 and Bash scripts.
  \item \textbf{Compilation:} C++20 toolchain (tested with GCC 11.4.0+) and CMake (tested with 3.22.1).
  \item \textbf{Benchmarks:} 57 workload traces from SPEC CPU2006~\cite{spec2006}, SPEC CPU2017~\cite{spec2017}, TPC~\cite{tpc}, MediaBench~\cite{fritts2009media}, and YCSB~\cite{ycsb}; {five PRADA microbenchmarks.}
  \item {\textbf{Runtime environment:} Linux (Ubuntu 22.04), Python 3.10.6, Slurm 21.08.5, and Podman 4.5.1 or Docker 24.0.5. Slurm is needed only for cluster runs.}
  \item {\textbf{Hardware:} x86-64 system; 32~GiB RAM for building and hardware evaluation; up to 4~GiB RAM per simulation. A cluster with at least 1000 cores is recommended for the full sweep.}
  \item {\textbf{Metrics:} IPC, weighted speedup, DRAM energy, memory access latency, and area and power.}
  \item {\textbf{Outputs:} PDF/PNG figures and text reports reproducing \figref{fig:xprada-perf}--\figref{fig:cgra}, \secref{sec:latency}, and \secref{sec:area-power-tradeoffs}.}
  \item {\textbf{Experiments:} (E1) performance and DRAM energy; (E2) hardware complexity.}
  \item {\textbf{How much disk space is required (approximately)?:} 80~GiB.}
  \item {\textbf{How much time is needed to prepare the workflow (approximately)?:} 75 minutes, largely unattended.}
  \item {\textbf{How much time is needed to complete experiments (approximately)?:} One day on a 1000-core cluster.}
  \item \textbf{Publicly available:} Yes.
  \item {\textbf{Automation:} Bash and Python scripts.}
  \item {\textbf{Archived:} Source DOI: \href{https://doi.org/10.5281/zenodo.21535719}{10.5281/zenodo.21535719}; traces DOI: \href{https://doi.org/10.5281/zenodo.21541650}{10.5281/zenodo.21541650}.}
\end{itemize}
}

\subsection{Description}

\subsubsection{Key results (R)}

\begin{enumerate}[label=(R\arabic*),widest=6,leftmargin=*,labelsep=0.5em,align=left]
  \item {Performance of four DRAM techniques and the composition of ChargeCache and MASA (\figref{fig:xprada-perf}--\figref{fig:xccmasa-perf}).}
  \item {DRAM energy overheads (\figref{fig:energy-overheads}).}
  \item {99th-percentile memory access latency (\secref{sec:latency}).}
  \item {Sensitivity to controller latency overhead (\figref{fig:latency-sensitivity}).}
  \item {Controller area and power overheads over custom implementations (\secref{sec:area-power-tradeoffs}).}
  \item {Performance of \titlename{} using programmable tables or reconfigurable logic (\figref{fig:cgra}).}
\end{enumerate}

\subsubsection{Experiment workflows (E)}

\begin{enumerate}[label=(E\arabic*),widest=2,leftmargin=*,labelsep=0.5em,align=left]
  \item {Performance and DRAM energy evaluation to reproduce R1--R4 and R6.}
  \item {Hardware complexity evaluation to reproduce R5.}
\end{enumerate}

\subsubsection{How to access}
The source code and scripts are available at \url{https://github.com/CMU-SAFARI/Terracotta} and traces are available on Zenodo at \url{https://doi.org/10.5281/zenodo.21541650}.

\subsubsection{Hardware dependencies}
The artifact runs on x86-64 systems with the following resources:
\begin{enumerate}
      \item {At least 32~GiB RAM for building the artifact and evaluating hardware complexity.}
      \item {Approximately 80~GiB of free disk space for source code, data, and results.}
      \item {Up to 4~GiB RAM and one core per simulation job.}
\end{enumerate}
{The full sweep comprises approximately 5000 jobs. We recommend a cluster with at least 1000 cores to complete it in approximately one day.}

\subsubsection{Software dependencies}
\begin{enumerate}
      \item {Linux (tested on Ubuntu 22.04).}
      \item {CMake (tested with 3.22.1).}
      \item {C++20 toolchain (tested with GCC 11.4.0+).}
      \item {Python 3.10, pip3, and the packages \texttt{matplotlib}, \texttt{pandas}, and \texttt{pyyaml}.}
      \item {Git, wget, tar, and unzip.}
      \item {Slurm 21.08.5 for cluster runs.}
      \item {Podman 4.5.1 or Docker 24.0.5 for containerized runs; required for E2 and optional for E1.}
\end{enumerate}
{The artifact's \path{README.md} provides installation instructions for common dependencies.}

\subsubsection{Workload memory traces}
We use workload memory traces collected from the SPEC CPU2006~\cite{spec2006}, SPEC CPU2017~\cite{spec2017}, TPC~\cite{tpc}, MediaBench~\cite{fritts2009media}, and YCSB~\cite{ycsb} suites.
{For PRADA, five microbenchmarks cover COPY, NOT, AND/OR, NAND/NOR, and XOR/XNOR computation operations in one bank and across 32 banks (\secref{sec:processing-using-dram}).}

\subsection{Installation}

{Clone the repository, then run the remaining commands from its root directory.}

\noindent\begin{minipage}{\columnwidth}
\textbf{{1. Obtain the source code.}}
\begin{cmdbox}
\begin{verbatim}
git clone \
    https://github.com/CMU-SAFARI/Terracotta.git
cd Terracotta
git submodule update --init --recursive \
    RTL/OpenROAD-flow-scripts
\end{verbatim}
\end{cmdbox}
\end{minipage}\par\smallskip

\noindent\begin{minipage}{\columnwidth}
\textbf{{2. Install Python dependencies.}}
\begin{cmdbox}
\begin{verbatim}
python3 -m pip install -r requirements.txt
\end{verbatim}
\end{cmdbox}
\end{minipage}\par\smallskip

\noindent\begin{minipage}{\columnwidth}
\textbf{{3. Download the traces.}}
\begin{cmdbox}
\begin{verbatim}
python3 download_traces.py
\end{verbatim}
\end{cmdbox}
\end{minipage}\par\smallskip

\noindent\textbf{{4. Build the tools.}}
{For E1, build Ramulator~2.0 on the host. We also support containerized builds. Please see the artifact's \path{README.md} for details.}
\begin{cmdbox}
\begin{verbatim}
./ramulator2/build_ramulator2_release.sh 
\end{verbatim}
\end{cmdbox}
\noindent{For E2, build the OpenROAD and Yosys container image. We only support a containerized build for E2.}
\begin{cmdbox}
\begin{verbatim}
./build_hw.sh
\end{verbatim}
\end{cmdbox}

\subsection{Experiment workflow}

\noindent\textbf{{E1: Performance and DRAM energy workflow.}}

\noindent\emph{1. Launch the sweep.} {We recommend using a Slurm cluster. The launcher throttles the number of simulation jobs (default 700), so please run the experiment launcher in detached mode so that experiments can continue after logout.}
\begin{cmdbox}
\begin{verbatim}
nohup ./run_with_slurm.sh all > sweep.log 2>&1 &
\end{verbatim}
\end{cmdbox}
\noindent{Alternatively, on a personal computer:}
\begin{cmdbox}
\begin{verbatim}
nohup ./run_with_personalcomputer.sh --no-container \
    all > sweep.log 2>&1 &
\end{verbatim}
\end{cmdbox}

\noindent\emph{2. Track simulation progress.} Due to the large number of jobs, we provide scripts that allow progress tracking.
\begin{cmdbox}
\begin{verbatim}
./check_progress.sh --watch  # live dashboard
./check_run_status.sh all    # completion status
\end{verbatim}
\end{cmdbox}

\noindent\emph{3. Parse and plot.} When all simulations are complete, run the following scripts to parse the results and reproduce the paper's key results.
\begin{cmdbox}
\begin{verbatim}
./parse_results.sh
./plot_all_figures.sh
\end{verbatim}
\end{cmdbox}
\noindent{Plots are written to \path{figures/}, CSV data to \path{results/csv/}, and text reports to \path{results/txt/}.}

\par\smallskip

\noindent\textbf{E2: Hardware complexity workflow.}

\noindent\emph{{Run synthesis.}} {This workflow can run on a personal computer:}
\begin{cmdbox}
\begin{verbatim}
nohup ./run_hw_with_personalcomputer.sh \
    > hw.log 2>&1 &
\end{verbatim}
\end{cmdbox}
\noindent On a Slurm cluster:
\begin{cmdbox}
\begin{verbatim}
nohup ./run_hw_with_slurm.sh > hw.log 2>&1 &
\end{verbatim}
\end{cmdbox}
\balance
\noindent{The CSV outputs are \path{synthesis_results.csv} and \path{terracotta_synthesis_results.csv} under \path{RTL/results/}. The summary is \path{RTL/results/txt/hw_complexity.txt}.}

\subsection{Evaluation and expected results}

\noindent\textbf{E1: Performance and DRAM energy.}

\noindent{E1 produces plots for R1, R2, R4, and R6 in \path{figures/}, in both PDF and PNG formats, and a text report for R3.}

\artifactresult{R1}{Performance}
Plots \texttt{figure9}--\texttt{figure13} reproduce \figref{fig:xprada-perf}--\figref{fig:xccmasa-perf}. They show that \titlename{} retains almost all of the {performance benefits of custom implementations}.

\artifactresult{R2}{DRAM energy}
Plot \texttt{figure14} reproduces \figref{fig:energy-overheads} and shows that \titlename{} incurs low DRAM energy overheads relative to custom implementations.

\artifactresult{R3}{Memory access latency}
\path{results/txt/latency_comparison.txt} reports the 99th-percentile memory access latency results in \secref{sec:latency}, showing a marginal 0.17\% increase over custom implementations.

\artifactresult{R4}{Controller latency sensitivity}
Plot \texttt{figure15} reproduces \figref{fig:latency-sensitivity} and shows the sensitivity of \titlename{} to controller latency overhead.

\artifactresult{R6}{Programmable tables and reconfigurable logic}
Plot \texttt{figure16} reproduces \figref{fig:cgra} and compares the performance of programmable table-based and CGRA-based \titlename{} implementations. Details on deriving the CGRA latencies are in \path{cgra/README.md}.

\par\smallskip
\noindent\textbf{E2: Hardware complexity.}

\noindent{E2 reproduces the controller area and power results (R5).}

\artifactresult{R5}{Controller area and power}
\path{RTL/results/txt/hw_complexity.txt} reports the results in \secref{sec:area-power-tradeoffs}. The additional controller area and power over custom implementations amount to 0.03\% of die area and 0.56\% of TDP of the reference Intel Xeon processor~\cite{intelemeraldrapids}.

\subsection{Experiment customization}
{The scripts support customization of}
(i)~the evaluated techniques,
(ii)~simulation parameters (e.g., number of instructions, number of concurrent jobs),
(iii)~the runtime environment (e.g., Slurm vs.\ personal computer),
(iv)~the workloads and mixes, and
(v)~containerization (e.g., Podman vs.\ Docker).
Details on how to customize the experiments are provided in the artifact's \texttt{README.md}.

\subsection{Methodology}
Submission, reviewing, and badging methodology:
\begin{itemize}\itemsep1pt
  \item \url{https://www.acm.org/publications/policies/artifact-review-and-badging-current}
  \item \url{https://cTuning.org/ae}
\end{itemize}

\end{document}